\documentclass[preprint,5p,times, twocolumn]{elsarticle}
\usepackage{hyperref}
\usepackage{graphicx}
\usepackage{epstopdf}
\usepackage{makecell}
\usepackage{amsmath}
\usepackage{float}
\usepackage{adjustbox}
\usepackage{rotating}
\usepackage{booktabs}

\usepackage{titlesec}

\titlespacing\subsection{0pt}{12pt plus 4pt minus 2pt}{0pt plus 2pt minus 2pt}
\titlespacing\subsubsection{0pt}{12pt plus 4pt minus 2pt}{0pt plus 2pt minus 2pt}

\usepackage{appendix}

\usepackage{algorithm}
\usepackage{algpseudocode}

\usepackage{amsmath}
\usepackage{breqn}

\usepackage{comment}

\usepackage{pifont}
\newcommand{\cmark}{\ding{51}}%
\newcommand{\xmark}{\ding{55}}%

\usepackage{amssymb}

\usepackage{multirow}

\journal{Journal Name}

\begin{document}

\begin{frontmatter}



\title{A Blockchain-based Decentralised and Dynamic Authorisation Scheme for the Internet of Things}



\author[]{Khizar Hameed \corref{mycorrespondingauthor}}
\ead{hameed.khizar@utas.edu.au}
\author[]{Ali Raza}
\ead{ali.raza@utas.edu.au}
\author[]{Saurabh Garg}
\ead{saurabh.garg@utas.edu.au}
\author[]{Muhammad Bilal Amin}
\ead{bilal.amin@utas.edu.au}

\address[]{Discipline of ICT, School of Technology, Environments, and Design,  University of Tasmania, Australia}

\cortext[mycorrespondingauthor]{Corresponding author}

\begin{abstract}
An authorisation has been recognised as an important security measure for preventing unauthorised access to critical resources, such as devices and data, within the Internet of Things (IoT) networks. To achieve authorisation, access control mechanisms are extensively utilised, restricting the user's actions within the network or system based on predetermined access control policies with specific control actions. Existing authorisation methods for the IoT network is based on traditional access control models, which have several drawbacks, including architecture centralisation, policy tampering, access rights validation, malicious third party policy assignment and control, and network-related overheads. The increasing trend of integrating Blockchain technology with IoT networks demonstrates its importance and potential to address the shortcomings of traditional IoT network authorisation mechanisms. However, existing Blockchain-based authorisation solutions for IoT networks overlook the importance of utilising the full potential of Blockchain technology and under-perform to handle the dynamicity of the underlying network in terms of malicious user behaviour, static policies, and auditability of user requests and resources. This paper proposes a decentralised secure, dynamic, and flexible authorisation scheme for IoT networks based on attribute-based access control (ABAC) fine-grained policies stored on a distributed immutable ledger. We design a Blockchain-based ABAC policy management framework divided into Attribute Management Authority (AMA) and Policy Management Authority (PMA) frameworks that use smart contract features to initialise, store, and manage attributes and policies on the Blockchain. To achieve flexibility and dynamicity in the authorisation process, we capture and utilise the environmental-related attributes in conjunction with the subject and object attributes of the ABAC model to define the policies. Furthermore, we designed the Blockchain-based Access Management Framework (AMF) to manage user requests to access IoT devices while maintaining the privacy and auditability of user requests and assigned policies. We implemented a prototype of our proposed scheme and executed it on the local Ethereum Blockchain. Finally, we demonstrated the applicability and flexibility of our proposed scheme for an IoT-based smart home scenario, taking into account deployment, execution and financial costs.

\end{abstract}

\begin{keyword}

Internet of things \sep Authorisation \sep Access control \sep Blockchain technology \sep Smart contracts \sep Dynamicity \sep Attribute-based access control


\end{keyword}

\end{frontmatter}



\section{Introduction}

Technological innovations for the Internet of Things (IoT) networks are constantly emerging, resulting in an explosion of connected resources such as sensors, embedded and intelligent devices \citep{atzori2010internet}. As a result, the globe is set to experience an unprecedented level of IoT in the following years, with 30 billion connected devices expected by 2025. While the expansion of IoT devices undoubtedly benefits the economy, it also creates severe security risks for IoT networks, as sensitive data from IoT devices is collected, analysed, aggregated, and shared across a variety of IoT-based platforms \citep{zafar2017trustworthy}. For instance, an adversary may exploit a system vulnerability by deploying a large number of untrusted devices in order to get unauthorised access to system resources. Data, services or applications, and hardware components (e.g., storage, computation) are the resources of IoT-based systems. Thus, protecting resources from unauthorised access is a challenging problem in an untrustworthy IoT-based environment \citep{yu2010achieving}.


Authentication and authorisation are the primary security requirements for protecting system resources. System administrators employ authorisation methods following authentication procedures to ensure that users are authentic and meet the initial security criteria for accessing specific resources \citep{hameed2021formally}. To achieve authorisation, access control acts as a protection mechanism to prevent an unauthenticated user from accessing sensitive resources by implementing regulation policies that specify what activities the users can undertake. For example, the regulation imposes selective access controls that govern who (an entity, such as a company) can access and what (a resource, such as data) can be accessed under particular scenarios \citep{sicari2015security}.


Numerous access control models have been adopted to protect system resources, and each model facilitates the mapping of user access levels to specified metrics. For example, Role-Based Access Control (RBAC) \citep{ferraiolo2003role} links user access levels to their roles in an organisation's structure, Attribute-based access control (ABAC) \citep{hu2013guide} links different access control levels to a user's attributes, and Capability-based Access Control (CapBAC) \citep{gusmeroli2013capability} assesses a user's capabilities to determine the appropriate access control level. Although many studies indicate that these traditional access control methods can be used in IoT-based environments, implementing them in real-world situations raises concerns due to architecture-level capabilities such as centralised networks, protocol constraints, device heterogeneity, scalability requirements, and functional limitations in IoT-based environments  \citep{pal2021blockchain}. Further, untrustworthy parties can potentially tamper with access policies in distributed environments, compromising system security by giving access unauthorised access to the users \citep{hussein2017community}. Apart from these integration concerns, each access control method has its own limitations, such as the rigour of fine-grained authorisation in RBAC, the management of attributes and policies in ABAC, and the flexibility of access privilege transfer in CapBAC. To summarise, implementing these traditional access control models to the IoT entails extensive research to overcome these challenges.

In recent years, the integration of Blockchain technology into the IoT has led to the development of robust and significant security solutions, most prominently for distributed and secure authorisation methods in IoT networks. Due to the numerous potential benefits of Blockchain technology, including decentralisation, immutability, consensus-mechanism, transparency, accessibility, autonomy, and accountability, a number of research studies are focusing on resolving a variety of issues associated with existing authorisation mechanisms for IoT devices, such as a single point of failure, data and policy tampering, insufficient control and transparency, and scalability. For instance, the initial Blockchain-based proposals \citep{dorri2017blockchain} \citep{maesa2017blockchain} \citep{pinno2017controlchain} \citep{andersen2019wave} for device authorisation in decentralised IoT environments are focused on resolving issues of a single point of failure and policy tampering by storing all access control policies as transactions on the distributed immutable ledger and enforcing all access control policies via the service providers. Furthermore, with the utilisation of smart contract features in designing the authorisation schemes for IoT, the authors in \citep{zhang2018smart} \citep{liu2020fabric} \citep{chai2021sfac} presented the access control schemes that exploit the Blockchain's smart contract capabilities to execute the access control policies automatically and to further achieve transparency by maintaining and storing access permissions on the different Blockchains such as public and private. Similarly, Hossein et al. \citep{shafagh2017towards} presented an access control mechanism for IoT that employs a distributed tempered-proof feature of Blockchain to store access rights. Furthermore, Novo \citep{8306880} presented a Blockchain-based distributed architecture for storing arbitrating roles and access control permissions on Blockchain to address the scalability problem of managing access to restricted devices in the IoT environments.

\subsection{Motivation and problem formulation}

Following a comprehensive study of the literature on approaches for secure authorisation in IoT-based networks, we identify the security requirements and objectives for designing a Blockchain-based trustworthy, flexible, fine-grained, and dynamic decentralised authorisation scheme. Then, based on the security requirements, objectives, and gaps in previous works, we derive motivation and formulate a research problem for our proposed scheme.

\subsubsection{Security requirements}

We outline the following security requirements that must be met in our proposed scheme to enable secure user authorisation to access critical resources (e.g., IoT devices) in IoT-based networks: (i) The proposed scheme must support the decentralised architecture in order to eliminate the single point of failure issue in existing centralised authorisation or access control mechanisms, where rules were created and managed by a single authority only, (ii) Critical resources, such as IoT devices and their stored data must be protected from unauthorised access, (iii) The scheme must ensure that only authorised or relevant authorities have the authority to define and modify policies, guaranteeing that policies are secure in a distributed tamper-proof manner, (iv) The privacy of users  who access system-critical resources and the security of their sensitive information must be ensured, (v) The user's requests to access system-critical resources must be auditable to establish their behaviour, which contributes to the dynamic nature of the deployed environment.

\subsubsection{Objectives}

Along with defining the security requirements, we outline the following objectives of our proposed Blockchain-based authorisation scheme, which must be met in order to provide users with secure access to IoT devices. For example, the proposed scheme must ensure a decentralised, secure, and automated process for capturing, storing, and managing information and policies on Blockchain. Second, the proposed scheme must determine the legitimate behaviour of network users and provide secure access to IoT critical resources. Third, the proposed scheme must provide dynamic authorisation by capturing context information (i.e., attributes) about network entities and their deployed environment and dynamically granting access to IoT resources in response to the changing behaviour. Fourth, the proposed scheme must enable fine-grained access control, allowing for granting or denying access to critical IoT resources based on numerous defined conditions or policies applied to a unique data resource. Fifth, the proposed scheme must be flexible to a broad range of IoT-based applications. Finally, the suggested approach must take into account the following parameters: storage, processing, and energy usage due to the resource-constrained nature of IoT devices.

\subsubsection{Research gap}

While prior research has demonstrated that Blockchain technology can be integrated into IoT environments to address the following issues with authorisation mechanisms such as single point of failure, access rights validation, data or policy tampering, lack of transparency, and accountability for granted resources, etc., however, these authorisation mechanisms have limitations that must be addressed in order to offer effective, secure, flexible, and dynamic authorisation of IoT devices by the use of prospective Blockchain features. For instance, the schemes \citep{dorri2017blockchain} \citep{maesa2017blockchain} \citep{andersen2019wave} solely rely on the Blockchain to store the expression of privileges to access resources in the form of transactions without fully utilising the Blockchain's capabilities, such as smart contracts. Moreover, to create and store predefined access rights on the Blockchain introduces issues with static access rights, such as in static authorisation schemes \citep{liu2020fabric} \citep{chai2021sfac} that are incapable of capturing the dynamic policies and coping with the dynamic nature of IoT networks, such as device mobility, behaviour patterns, the need to perform critical operations, device compromise and failure, and so on. Some methods rely on traditional access control models such as access control list (ACL) \citep{zhang2018smart} and RBAC \citep{novo2018blockchain} \citep{xu2021distributed} \citep{andersen2017wave} \citep{cruz2018rbac}, respectively, and each of which has its own set of limitations in terms of their inflexible nature, time consumption, error-prone and coarse-grained level. Further, the authorisation schemes based on CapBAC models \citep{chai2021sfac}

\citep{xu2018blendcac} \citep{nakamura2019capability} \citep{xu2018blendcac2} \citep{xu2019exploration} \citep{nakamura2020exploiting} \citep{chai2021bhe} grant access to resources solely based on capability tokens created for each policy. Further, employing CapBAC models in authorisation processes is a time-consuming task requiring generating and distributing tokens (or capabilities) to all subjects for each resource. It is also possible that selecting a specific capability while making a request overburdens the network, which significantly impacts system performance when there are many users. Finally, the authorisation methods based on ABAC models have limitations in terms of achieving the privacy of the attributes \citep{liu2020fabric} \citep{wang2019attribute} \citep{sultana2020data} \citep{zhang2020attribute} \citep{qin2021lbac}, collecting and managing the large number of policies and achieving the flexibility and fine-grained level of access rights \citep{wang2019attribute} \citep{sultana2020data} \citep{qin2021lbac} .


Taking into account the inherent issues with traditional access control models and their integration with IoT, and in order to address the limitations of existing Blockchain-based authorisation schemes, we proposed a Blockchain-based trustworthy, flexible, fine-grained, and dynamic decentralised authorisation mechanism for IoT that makes use of smart contracts to ensure the full capability of the Blockchain. Furthermore, we use the ABAC model to build fine-grained rules consisting of a subject, object, and environment-related attributes to ensure the flexibility and dynamic nature of our proposed mechanism. To manage fine-grained policies efficiently, we developed a policy-management framework composed of various smart contract-based modules that ensure the secure provision of the required access policies to users requesting to access IoT devices. For the most part, our solution idea is centred on using smart contracts to implement attribute-based Access Control regulations. Smart contracts running on the blockchain control both the policies and the attributes required for evaluating them. To demonstrate the practicality of our proposed mechanism, we create a proof-of-concept prototype comprised of a local private Ethereum Blockchain that executes the smart contract functionality and logic.

\subsection{Contributions}

The following constitute the primary contributions to this paper:

\begin{itemize}
    \item  We propose a Blockchain-based decentralised, secure and flexible authorisation scheme for IoT networks, featured with the smart contracts-based ABAC model to enforce the execution of authorisation process for providing users with secure access to IoT resources based on dynamic and fine-grained policies stored on the distributed immutable ledger.

    \item We designed the ABAC-PMF with smart contract features, consisting of two sub-components: AMA and PMA, which handle attributes and policies on the Blockchain, respectively. The AMA initialises, stores, and manages subject, object, and environment-related attributes to enable flexible and fine-grained access permissions. On the other hand, PMA initialises, stores, and manages policies to offer dynamic access to users based on environmental attributes and defined actions.

    \item We designed the AMF by using the features of smart contracts with the aim of securely managing user requests to access IoT devices based policies. Furthermore, AMF maintains a look-up table to ensure the privacy of user information and the auditability of user requests and assigned policies.

    \item We designed a prototype of our proposed scheme as a proof-of-implementation and executed the functionality of smart contracts on the local Ethereum Blockchain setup. Furthermore, to demonstrate the practicality and applicability of our proposed scheme, we computed the deployment and execution costs and the financial cost of enabling secure access to IoT devices via an IoT-based smart home scenario.

\end{itemize}

\subsection{Paper organisation}

The organisation of this paper is structured as follows: Section \ref{overview} covers the background to our work and literature review of existing Blockchain-based authorisation schemes. A detailed system architecture of our proposed scheme consisting of the network model and their assumptions is presented in the section \ref{systemarchitecture}.  Section \ref{proposedauthorisation} provide a detailed description of the authorisation process, proposed algorithms, execution flow and qualitative security analysis. The implementation and evaluation framework of our proposed scheme is presented in section \ref{evaluation}. The performance analysis of the proposed scheme in terms of deployment and execution costs followed by the financial cost of deploying IoT-based smart home scenario is discussed in the section \ref{performance}. Finally, we conclude our paper in section \ref{conclusion}.

\section{Background and related work} \label{overview}

This section begins with a brief discussion of some fundamental concepts, followed by an introduction to security solutions proposed in recent literature that utilise Blockchain technology for providing secure access to users using numerous authorisation mechanisms.


\begin{table*}[!t]
\centering
\scriptsize
\caption{A comparative analysis of our proposed scheme with existing smart contract based authorisation or access control mechanisms for IoT}
\begin{tabular}{|p{1.5cm}|p{1.5cm}|p{1.5cm}|p{1.5cm}|p{0.5cm}|p{0.5cm}|p{0.5cm}|p{0.5cm}|p{0.5cm}|p{0.5cm}|p{0.5cm}|}
\hline
\multirow{2}{*}{\textbf{Ref}} & \multirow{2}{*}{\textbf{\parbox{2cm}{Access control \\ model}}} & \multirow{2}{*}{\textbf{\parbox{2cm}{Network \\type}}} & \multirow{2}{*}{\textbf{\parbox{2cm}{Storage \\ requirement}}} & \multicolumn{7}{c|}{\textbf{Achieved security requirements and objectives}}                                        \\ \cline{5-11}

                           &                            &                            &                            & \textbf{DD} & \textbf{RP} & \textbf{P} &\textbf{FL} & \textbf{DN} & \textbf{FG} & \textbf{AD} \\ \hline

                           \citep{ouaddah2016fairaccess}& A generic distributed access control framework&N/A&Single&\cmark&\xmark&\cmark&\xmark&\cmark&\cmark& \xmark\\ \hline

                           \citep{dukkipati2018decentralized}  &     A generalised access control method itilising XACML policies                &            Private              &          Single                  &        \cmark   &   \xmark        &      \xmark     &     \xmark      &     \xmark      &  \xmark     & \xmark   \\ \hline

                           \citep{ali2020xdbauth}    &        A generic decentralised blockchain based access control framework               &   Private                       &    Multiple                        &       \cmark    &    \cmark       &       \cmark    & \xmark   &       \xmark     &   \xmark    &  \xmark   \\ \hline

                           \citep{esposito2021blockchain}&A generalised approach based on XACML Policies&Private&Single&\cmark&\xmark&\xmark&\cmark&\xmark&\xmark&\xmark \\ \hline

                         \citep{tan2021blockchain}  & A generalised approach based on decentralised identifiers&N/A&Multiple&\cmark&\xmark&\cmark&\xmark&\xmark&\cmark&\xmark \\ \hline

                          \citep{zhang2018smart}&ACL&Private&Single&\cmark&\xmark&\cmark&\xmark&\cmark&\xmark& \xmark \\ \hline

                          \citep{andersen2017wave} &RBAC&Private&Single&\cmark&\xmark&\cmark&\xmark&\xmark&\cmark&\xmark \\ \hline

                   \citep{cruz2018rbac}        &        RBAC                    & Public                           &      Single                      & \cmark          &       \cmark    &      \xmark     &        \xmark   &     \xmark      &    \xmark     & \xmark  \\ \hline

                          \citep{8306880}&RBAC&Private&Single&\cmark&\cmark&\xmark&\cmark&\cmark&\xmark& \xmark \\ \hline

                           \citep{xu2021distributed}&RBAC&Consortium&Multiple&\cmark&\cmark&\cmark&\xmark&\cmark&\xmark&\xmark \\ \hline

                    \citep{xu2018blendcac}       &        CapBAC                    &           Private                 &      Multiple                      &        \cmark   &    \xmark       & \xmark &    \xmark     &      \xmark     &      \cmark     & \xmark           \\ \hline

                     \citep{nakamura2019capability}      &        CapBAC                    &            Private                &    Single                        &      \cmark     &      \xmark     &     \xmark      &      \cmark     &      \xmark     &  \cmark &   \xmark     \\ \hline

                         \citep{xu2018blendcac2}  &           CapBAC                    &           Private                 &      Multiple                      &     \cmark      &    \xmark       & \xmark &    \xmark     &      \xmark     &      \cmark     & \xmark           \\ \hline

                     \citep{xu2019exploration}      &         CapBAC                   &         Private                   &           Multiple                 &       \cmark    &         \xmark  &     \xmark      &       \xmark    &       \xmark    &   \cmark    &   \xmark \\ \hline

                   \citep{nakamura2020exploiting}     &       CapBAC                  &     Private                     &        Single                    &        \cmark   &      \xmark     &        \xmark   &      \cmark     &        \xmark   &     \cmark  &   \xmark \\ \hline

                  \citep{chai2021sfac} &CapBAC&N/A&Multiple&\cmark&\xmark&\xmark&\xmark&\xmark&\cmark&\xmark \\ \hline
                   
                   \citep{chai2021bhe}&CapBAC&N/A&Multiple&\cmark&\xmark&\cmark&\xmark&\xmark&\xmark& \xmark \\ \hline

                     \citep{wang2019attribute}      &         ABAC                   &          Public                  &     Single                       &     \cmark      &       \xmark  &\xmark  &    \cmark       &        \xmark   &    \cmark       &  \xmark         \\ \hline
                           
                       \citep{yutaka2019using}  &   ABAC                         &    Private                     & Single   & \cmark   &   \xmark                     &    \cmark       &      \cmark      &         \cmark   &     \cmark       &      \xmark              \\ \hline

                      \citep{liu2020fabric} &ABAC&Private&Single&\cmark&\xmark&\xmark&\xmark&\cmark&\cmark&\xmark \\ \hline

                     \citep{sultana2020data} &ABAC&Private&Single&\cmark&\xmark&\xmark&\xmark&\xmark&\xmark&\xmark \\ \hline

                            \citep{zhang2020attribute}     &       ABAC                  &      Private                    &      Single                      &      \cmark     &   \xmark        &     \xmark      &       \cmark    &       \cmark    &   \cmark    &  \xmark  \\ \hline

                       \citep{putra2021trust}     &         ABAC                &  Public                        &    Multiple                        &         \cmark  &   \cmark        &       \cmark    &       \cmark &        \cmark   &  \cmark    & \xmark    \\ \hline

                      \citep{qin2021lbac} &ABAC&N/A&Multiple&\cmark&\cmark&\xmark&\xmark&\cmark&\xmark& \xmark \\ \hline

                     Our Proposed Scheme  &ABAC&Private&Multiple&\cmark&\cmark&\cmark&\cmark&\cmark&\cmark&\cmark \\ \hline

\end{tabular}

{\raggedright \textbf{DD} =  Decentralised and Distributed, \textbf{RP} =  Resource Protection, \textbf{P} = Privacy, \textbf{FL}= Flexibility, \textbf{DN} = Dynamicity,  \textbf{FG} = Fine Grained, \textbf{AD} = Auditability  }
\label{Tab:comparison}
\end{table*}

\subsection{Background}

This section discusses Blockchain technology and the ABAC concept, which serve as the background to our work.

\subsubsection{Blockchain technology}

The Blockchain is the primary technology that underpins Bitcoin and other cryptocurrencies. Blockchain technology allows for the development of a trusted network of untrusted nodes in which all network nodes must validate transactions. In Blockchain, nodes keep track of and verify new transactions in blocks, eliminating the need for centralised intermediaries. Due to the fact that all nodes maintain a history of transactions in the form of linked hashes that make up the Blockchain, any changes made to previously-stored transactions will be detected by the Blockchain. Miner nodes accept and verify new transactions prior to insertion into the Blockchain via the consensus mechanism.  In most consensus mechanisms, miner nodes must solve puzzles based on their computational resources to add verified blocks to the Blockchain.

Another significant feature of Blockchain technology is implementing smart contracts, which are small pieces of code written in the Blockchain and activated when certain conditions are met. The successful implementation of this feature in several Blockchain-based platforms, including Ethereum \citep{wood2014ethereum}, enables the development of decentralised and trusted execution platforms, also known as decentralised applications (dApps).  Instead of simply recording data on an immutable ledger, smart contracts strive to enhance the capabilities of Blockchain by managing complicated and autonomous computations. The key principle behind the implementation is to put executable codes on the Blockchain and make peers execute them by coordinating and cooperating with each other. Further, the peer nodes establish an agreement when evaluating the code and storing the results in the Blockchain.


The emergence of outstanding features of Blockchain technology in IoT led to the design and implementation of a robust platform or applications for transparently and trustlessly transferring information across the network. For example, the decentralised and distributed nature of Blockchain enables IoT components to join the network without the involvement of a third party and communicate with one another via secure cryptographic protocols, thereby ensuring the scalability, security, and reliability of both the component and data. Further, the data is stored on the immutable ledger as permanent transactions, enabling the network nodes to verify the transactions in the event of a discrepancy, hence introducing a high level of auditability. In terms of smart contracts, automatic execution of certain conditions is possible to aid with a wide range of functions such as controlling access permissions and executing access requests in secure decentralised authorisation systems.

\subsubsection{Attribute-based access control (ABAC)}

ABAC is an authorisation paradigm that determines user access to system resources based on attributes (or characteristics) rather than roles. The objective of ABAC, like other access models, is to secure both critical system resources and sensitive data from unauthorised users and their malicious actions. ABAC ensures that access decisions are made based on the attributes of the subject (i.e., user), object (i.e., resources), and environment involved in an access event.  For example, the ABAC model determines that users who do not have "authorised" attributes as determined by an organisation's security policies will not be permitted access to resources or data.

In ABAC, the access policies are described as a logical arrangement of attributes and actions used to provide access to critical resources. In an ABAC model, policies make use of a variety of attributes in conjunction with Boolean logic formulae to determine who is making the request, what resource is being used, and what action should be taken. There are two types of attribute values: set and atomic. The difference between atomic and set-valued characteristics is that set-valued attributes hold more than just one atomic value.

\subsection{Related work}

Blockchain technology has earned a competitive edge in IoT networks by virtue of its multiple features, including decentralisation, distributed security, tamper-resistant, transparency, and autonomy. Recent research has demonstrated the efforts being made to incorporate these features into the IoT in order to develop more robust and efficient solutions that address some of the issues associated with traditional authorisation or access control methods, including single point of failure, the untrustworthy delegation of access rights, third-party validation, modification of access rights, lack of transparency and control, and scalability.

\subsubsection{Simple Blockchain-based approaches}

This section discusses a few Blockchain-based authorisation mechanisms for IoT networks that rely on the Blockchain as an underlying architecture (without smart contracts) for defining and storing access policies.

For instance, Dorri et al. \citep{dorri2017blockchain} presented a Blockchain-based mechanism for resolving the issue of access control for multiple interacting entities in an IoT-based decentralised smart home environment, such as the cloud, users, and IoT devices. Each home has its own private Blockchain in this scheme, where the multiple miners keep a policy header to control all home access requests. However, this mechanism mainly focused on distribution and recording access privileges on the immutable ledger in Blockchain. Further, Maesa et al. \citep{maesa2017blockchain} proposed a blockchain-based access control delegation mechanism for storing and transferring expressions about access rights to a given resource in the form of transactions. To secure the validation of access rights, Andersen et al. \citep{andersen2019wave} proposed a scalable decentralisation authorisation mechanism that grants access to users via advanced cryptographic features. In addition to access delegation, a reverse discoverable decryption protocol protects shared access privileges between administrative domains. Pinno et al. \citep{pinno2017controlchain} presented a distributed architecture built on Blockchain technology to govern access to IoT devices and ensure transparent authorisation. This architecture secures relationships between users and devices using different access control models, including ACL, CapBAC, and ABAC. Also, Ding et al. \citep{ding2019novel} proposed an IoT access management system built on Blockchain that simplifies the process of authorising IoT devices based on their attributes. The immutability feature of Blockchain is used to store attribute distribution in the form of transactions for permission and revocation. While these access control solutions addressed the aforementioned issues effectively by maintaining access policies on a tamper-proof ledger, they vastly overstated the full potential of blockchain technology by utilising its advanced features for performing various computation functions.

\subsubsection{Smart contracts-based approaches}

As the primary objective of this paper is to propose a Blockchain-based decentralised, secure authorisation scheme for IoT networks through the use of Blockchain's smart contracts feature, we provide a detailed overview of authorisation mechanisms for IoT networks that make use of smart contracts and support for various access control models. Furthermore, we perform a detailed comparison of existing smart contract-based schemes to our proposed scheme in terms of security requirements and objectives fulfilled, as presented in Table \ref{Tab:comparison}.

Along with access delegation, a reverse discoverable decryption protocol is utilised to maintain the privacy of access rights shared between multiple administrative domains. Pinno et al. \citep{pinno2017controlchain} proposed a distributed architecture built on Blockchain technology to manage access to IoT devices and ensure a transparent authorisation of IoT devices. By utilising multiple access control models such as ACL, CapBAC, and ABAC, this architecture enables the secure establishment of communication between individuals and devices while mapping their unique characteristics to the relationships utilised in access control authorisation. Same with the previous solutions, Ding et al. \citep{ding2019novel} presented a Blockchain-based access management system for IoT systems that simplifies the process of authorising IoT devices based on their attributes. In this system, the immutable feature of Blockchain technology is utilised to store the distribution of attributes in the form of transactions for authorisation and revocation processes. While these existing solutions address the single point of failure and data or policy tempering issues inherent in existing authorisation or access control schemes by storing access policies on a tamper-proof ledger, they frequently overlook the full potential of Blockchain.

To broaden the functional scope of blockchain technology through smart contracts (i.e., executable codes stored on the Blockchain), the Blockchain has been hailed as a leading framework for developing decentralised and trustworthy applications, attracting significant interest from researchers in the IoT community. For example, Zhang et al. \citep{zhang2018smart} propose a smart contract-based access control mechanism to overcome the limitations of prior works that exclusively used the Blockchain to store access rights. This mechanism provides decentralised and trustworthy access control for IoT systems using smart contracts, comprising access control, judge, and register contracts. However, this mechanism relied on a static access control list, which limited its usage to determining and revoking user access privileges on all objects. As a result, this mechanism lacks efficiency and scalability.

Andersen et al. \citep{andersen2017wave} presented the WAVE, a decentralised authorisation scheme, with RBAC to enable the secure authorisation of IoT devices without the need for a central trusted party. The delegation of resources on the public Blockchain is accomplished through smart contracts in conjunction with the delegation trust mechanism. Cruz et al. \citep{cruz2018rbac} proposed a Blockchain-based access roles mechanism for the secure interaction of several trans-organisations through the use of smart contracts. Likewise, this technique used RBAC to create user roles and assignments at the organisational level and verify a user's role ownership. Further, Novo \citep{8306880} presented a Blockchain-based distributed architecture for storing arbitrating roles (i.e., RBAC) and access control permissions on Blockchain to address the scalability problem of managing access to restricted devices in the IoT. Numerous restricted IoT devices are linked concurrently by various flexible management hubs distributed across the blockchain network. The proposed architecture uses smart contracts to create the manager node that manages permissioned Blockchain access control rights. However, this method also utilised the RBAC. Xu et al. \citep{xu2021distributed} proposed a Blockchain-based decentralised authorise and permission management method for healthcare data accessed by different entities related to healthcare such as hospitals, physicians and medical insurance companies. Authorisation information contains URLs (Universal Resource Locators) to identify the data approach especially and decouple patient privacy data from the Blockchain to ensure the effectiveness of the data processing. In this method, the RBAC model is utilised to define the authorisation rules in smart contracts, allowing for executing, recording, and tamper-proofing all private data authorisation operations. However, the limitations of RBAC-based authorisation methods are that users are granted access privileges based on their roles, and all users with similar roles have access to the same amount of data regardless of their restricted authorisation to fine-granularity access rights. Additionally, because RBAC models are static in defining access policies and limiting mechanisms to only achieving coarse-grain access levels, utilising such models in authorisation mechanisms can be time-consuming for defining and accessing policies and error-prone when the system has a large number of roles.

CapBAC is another access control model used by various IoT network authorisation methods to ensure that users have secure and efficient access to resources. Additionally, this model is intended to overcome the limitations of RBAC by giving access via identity tokens rather than the users' roles. To demonstrate the concept of CapBAC in an IoT-based authorisation mechanism, Xu et al. \citep{xu2018blendcac} presented a decentralised CapBAC framework, called BlendCAC, that leverages smart contracts and blockchain technologies to enable safe access to IoT devices. An identity-based capability token management method is suggested in this framework, which makes use of smart contracts for the initialisation, registration, transmission, and revocation of access rights on Blockchain. Furthermore, the work \citep{xu2018blendcac2} proposes an extension of the work \citep{xu2018blendcac} by employing a similar approach but focusing on additional performance metrics such as block generation time.
Furthermore, the same approach \citep{xu2018blendcac} was used in \citep{xu2019exploration} to investigate the CapBAC model in space situation awareness (SSA) applications. Space situation awareness (SSA) entails the surveillance of active and inactive local space objects and the characterisation of the space environment via the collection and processing of sensor data. Nakamura et al. \citep{nakamura2019capability} propose another authorisation method for IoT networks based on the CapBAC model, in which smart contracts are primarily used to store and maintain capability tokes for giving access to IoT devices. Tokens are generated as units of action performed by users in this method, and the delegation graph is used to map users to resources. The work \citep{nakamura2020exploiting} is an extension of the \citep{nakamura2019capability}, with an emphasis on the implementation of functions (i.e., capability, delegation graph) proposed in the earlier work for evaluating and analysing gas use.
Moreover, Chai et al. \citep{chai2021sfac} presented an access control method based on CapBAC for IoT networks, called SFAC, that incorporates smart contracts and Blockchain technology to ensure that only authorised users have access to critical information. The suggested method made use of a tokens mechanism that enables users to request numerous resources in batches rather than individually. An extension of \citep{chai2021sfac} is presented in \citep{chai2021bhe}, with a particular emphasis on the analysis of the results. However, the disadvantage of CapBAC-based authorisation systems or access control models is that users are granted access to resources solely based on capability tokens created for each policy. Further, employing CapBAC models in authorisation processes is a time-consuming task that requires generating and distributing tokens (or capabilities) to all subjects for each resource. It is also possible that selecting a specific capability while making a request overburdens the network, which significantly impacts system performance when there are many users.

To overcome the limitations of RBAC and CapBAC models in terms of nature of access rights (i.e., static), assignment level (i.e., coarse-grained), reduce the computation cost of generating tokens and provide dynamic authorisation of IoT devices, ABAC models are frequently deployed. In ABAC models, different attributes are used to grant access privileges to IoT devices. By utilising the device attributes of the ABAC models in IoT-based networks to provide secure access to underlying resources, Yutaka et al. \citep{yutaka2019using} proposed a Blockchain-based access control scheme by utilising the smart contracts to write and enforce the policies. In this scheme, a range of smart contracts, from registration of attributes to defining the ABAC policies, is written and executed on Ethereum Blockchain. Finally, the proposed scheme is deployed on the Local Ethereum Blockchain to demonstrate its feasibility and evaluate its monetary cost. An extension of the previous work \citep{yutaka2019using} is presented in the \citep{zhang2020attribute}, with a focus on enhancing performance evaluation parameters such as deployment and operating cost. Wang et al. \citep{wang2019attribute} proposed a Blockchain-based distributed access control framework for IoT networks by utilising the subject and objects based attributes. In this framework, attributes such as manufacturer and object specifications provide a finer-grained degree of access control for resource-constrained IoT devices. Liu et al. \citep{liu2020fabric} introduced Fabric-IoT, a fabric-based access control system that leverages the ABAC and Hyperledger fabric Blockchain architectures to enable dynamic and fine-grained access control rights for IoT networks. Three smart contracts with distinct functionality (device, policy, and access) are created and deployed on the Blockchain in this scheme to ensure the secure execution of access rights. For access rights and delegation of access rights in IoT based networks, Ali et al. \citep{ali2020xdbauth} proposed a decentralised Blockchain-based architecture called xDBAuth, in which the smart contracts are written and deployed hierarchically, providing permission assignment and access control for both local and global devices, as well as authenticating in their parent IoT domains. A similar approach to \citep{liu2020fabric} is given in \citep{sultana2020data}, with the goal of providing data sharing and access control across IoT devices via the use of Blockchain and smart contracts. Qin et al. \citep{qin2021lbac} utilised an ABAC model to propose a lightweight blockchain-based access control scheme called LBAC. Blockchain is used instead of untrusted cloud servers to perform the outsourcing decryption on attributes determining the secure access to resources. To achieve the dynamic nature of authorisation mechanisms for IoT networks, Putra et al. \citep{putra2021trust} presented a decentralised access control mechanism based on the ABAC model in conjunction with a trust and reputation system. The proposed mechanism sought to achieve dynamic access privileges by focusing on IoT device compromise scenarios.

Furthermore, various Blockchain-based generic authorisation or access control systems have been proposed that utilise the smart contract feature to secure user access to IoT-based networks. Dukkipati et al. \citep{dukkipati2018decentralized} proposed a general access control mechanism based on Blockchain technology to ensure the security and privacy of IoT access resources. This approach makes use of the smart contract feature to establish the structure of XACML (Extensible Access Control Markup Language) based policies that can be executed on the Blockchain. Further, Ouaddah et al. \citep{ouaddah2016fairaccess} presented a decentralised reference model based on the Hyperledger Fabric Blockchain for access control management tasks such as issuing, fetching and assigning rights to users via new types of transactions created by modifying the basic transaction structure of Hyperledger Fabric Blockchain. Esposito et al. \citep{esposito2021blockchain} designed and implemented a blockchain-based approach for authentication and authorisation in smart cities, which is further integrated with FIWARE to overcome its centralised architecture limitation. In addition, the proposed approach defined the structure of policies to be recorded on the Blockchain using XACML. Tan et al. \citep{tan2021blockchain} proposed a generic Blockchain-based authorisation, access control and revocation method for Green Smart Devices (GSDs) to enable safe access to Green IoT (GIoT) devices through the use of decentralised identities. By utilising smart contracts implemented on the Blockchain, decentralised IDs are issued, managed, and revoked.

\begin{figure*}[t]
    \centering
    \includegraphics[width=14cm,height=8cm]{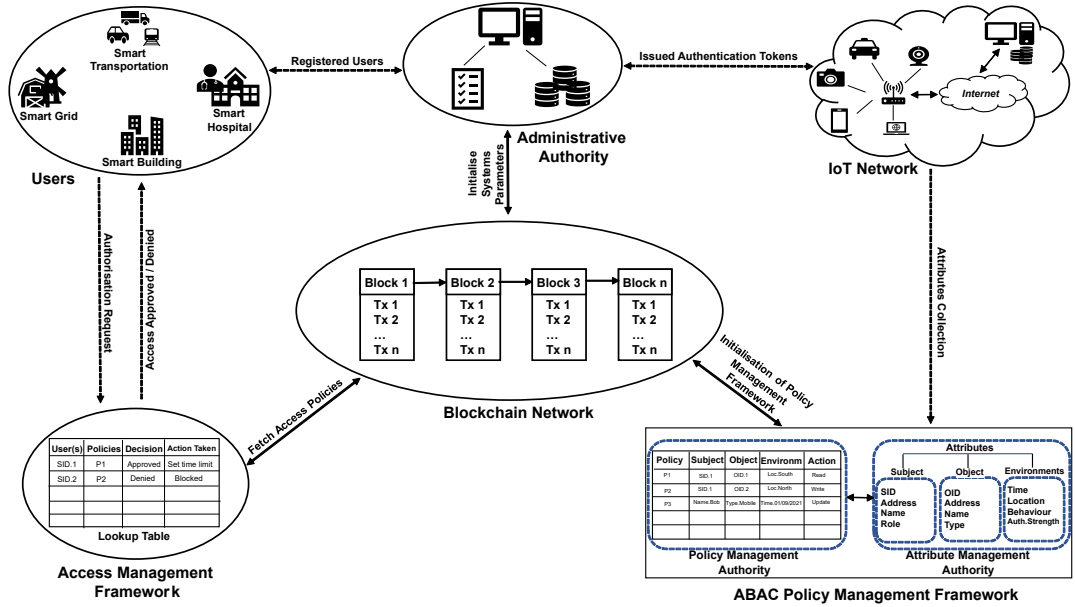}
    \caption{Network model}
    \label{networkmodel}
\end{figure*}

\section{System architecture} \label{systemarchitecture}

This section describes the system architecture for our proposed decentralised, flexible and dynamic authorisation scheme for IoT devices based on Blockchain technology. To simplify and clarify the system architecture, we first divide it into network model components and then define the underlying assumptions for each component.

\subsection{Network model}\label{sec:networkmodel}
Fig. \ref{networkmodel} illustrates the network model for our proposed Blockchain-based secure, flexible and dynamic decentralised authorisation mechanism for IoT devices. Our network model comprises the following primary interacting components that enable authorised users to securely interact with IoT devices via the access management framework in conjunction with the Ethereum-based ABAC policy framework, which defines access policies using subject, object, and environment-related attributes. The details for each component are listed below.

\subsubsection{Administrator authority}
The administrative authority (AA) is a critical component of the network model, and it is responsible for the following activities:

\begin{itemize}
    \item Firstly, it is responsible for initialising the overall system configuration, which includes generating public parameters and uploading them to the Blockchain. Since our proposed mechanism is based on the Ethereum Blockchain, this component assigns Ethereum addresses \textit{EA} = \{$ea_{1}$, $ea_{2}$, $ea_{3}$, \dots, $ea_{N}$\} and public keys \textit{PK} = \{$pk_{1}$, $pk_{2}$, $pk_{3}$, \dots, $pk_{N}$\}  to all interacting modules in this scenario.
    
    \item Secondly, it is responsible for issuing authentication tokens to IoT devices based on the inherent attributes of both users (subjects) and IoT devices (object attributes). We have a \textit{N} number of IoT devices in our work, and hence tokens are represented as $T$ = \{$T^1_{SO}$, $T^2_{SO}$, $T^3_{SO}$, \dots, $T^N_{SO}$\}. Our previous work \citep{hameed2021formally} describes the whole process of creating and issuing Blockchain-based authentication tokens in a decentralised IoT network.
    
    \item Finally, this component is responsible for registering environmental attributes (also known as context attributes) associated with both subjects and objects to define access control policies in the policy management framework. Our previous work \citep{hameed2021context} details the process of collecting, characterising and controlling environmental attributes in IoT networks.
\end{itemize}

\subsubsection{ABAC policy management framework (ABAC-PMF)}

The ABAC-PMF is a critical component of our proposed network model. This component serves as the full node in the Blockchain network, as it is responsible for the overall process of collecting, storing, and managing the attributes essential to define access control policies. This component is further subdivided into two components, the details of which are as follows:

\begin{itemize}
    \item \textbf{Attributes management authority (AMA):} The AMA is solely responsible for collecting, storing, and managing attributes on the Blockchain. We classified the attributes into three categories for the ABAC model: subject attributes (SA), object attributes (OA), and environmental attributes (EA). Then, for each attribute type, we create a smart contract whose primary role is to collect, store, and manage the attributes associated with each category on the Blockchain and interact with other components in the network model. We designated smart contracts as $S_{SA}$, $S_{OA}$, $S_{EA}$, respectively, for each attribute type such as subject, object, and environment. The section \ref{regisandmanageattributes} goes into detail about each smart contract and its working mechanism, including functions and logic.
    
    \item \textbf{Policy management authority (PMA):} The PMA is tasked with storing and managing access control policies, defined in terms of stored attributes, to the Blockchain. Similarly to creating smart contracts for attribute types, we created and deployed another smart contract named with policy management authority, denoted as $S_{PMA}$. The $S_{PMA}$ is responsible for defining, storing, and managing access policies and providing secure communication with the access management authority.

\end{itemize}

\subsubsection{Access management framework (AMF)}

As with the ABAC-PMF component, the AMF is a critical component of our network model and acts as the full node in the Blockchain network. It is responsible for handling users' access requests to IoT devices in accordance with the access policies defined in the ABAC-PMF. The technique by which user access requests are handled is a mapping procedure that establishes the relationship between the subject and the object following stated policies. To manage user access requests, we designed a lookup table that categorises user requests according to subject attributes, object attributes, and policies and determines whether users are permitted to access resources or not. Additionally, we built an action taken control mechanism within the access management framework to identify the following steps to respond to user access requests. To realise this concept, we created a smart contract $S_{AMF}$ that manages user requests for accessing resources.

\subsubsection{Blockchain network}

In our proposed network model, the administrative authority is responsible for initialising and configuring the Blockchain network parameters to connect the other components, such as the ABAC-PMF and AMF. In our case, we deployed a single Blockchain as a public Blockchain, namely with \textit{PB}, which provides a decentralised environment for the flexible and dynamic authorisation of IoT devices through the execution of access control policies defined in Ethereum-based smart contracts. Additionally, it enables secure administration and storage of access control policies on a distributed tamper-resistant ledger. A PB is operated by individual miners that seek to earn monetary rewards through block mining.

\subsubsection{Users}

The users are authorised individuals who possesses specific unique characteristics (i.e., subject attributes), as such \textit{Users} = \{$U^1_s$, $U^2_s$, $U^3_s$, \dots, $U^N_s$\}, and is granted access to IoT devices (i.e., resources) within an IoT network based on defined access policies for executing various actions. For instance, in a smart building setup, users are assigned their unique access policies to control and manage their resources, such as door locks, lights, and security alarms for each apartment. As our proposed Blockchain-based authorised mechanism is intended to be general-purpose with multiple benefits such as trustworthiness, flexibility, and dynamicity, it can be applied to any IoT-based environment, including smart hospitals, smart transportation, and smart buildings and smart grids. Users act as light nodes in our network model since they communicate with the access management authority to gain access to IoT devices.

\subsubsection{IoT devices}

In our network model, IoT devices act as a collection of resources with some distinct characteristics (i.e., object attributes) and can be accessed only by authorised users to accomplish a series of tasks within an IoT network. In our scenario, we have a \textit{N} number of IoT devices with object attributes, denoted by \textit{Devices} = \{$D^1_o$, $D^2_o$, $D^3_o$, \dots, $D^N_o$\}. However, the main aim of an adversary is always to get access to these IoT devices, whether single or multiple, to gain unauthorised control of the entire network and perform various malicious actions such as device failure, data manipulation, and denial of service attacks. IoT devices serve as light blockchain nodes in our network model, connecting to the ABAC policy management framework.

\subsection{Network model assumptions}

Our proposed scheme aims to achieve decentralised, secure, fine-grained and dynamic authorisation of users to access IoT devices through an ABAC model implemented using Blockchain technology. Our network model is made up of several interconnected components that work and communicate together to realise the concept of IoT device authorisation in a decentralised manner. We made the following assumptions for each interacting component:

\begin{itemize}
    \item The AA component has the full authority to initialise, control and manage all the other components of the network model.  Also, it distributes the system parameters such as Ethereum addresses, public and private keys to all other components in a secure way. 
    
    \item For the IoT network, we used the attacker model proposed in our previous work \citep{hameed2021formally} to define the attackers' capabilities. By employing this attacker model, miners act as completely trusted resource providers, ensuring that IoT resources are authentic and trustworthy. 

    \item The miner nodes control and manage the ABAC-PMF, which act as the full nodes and communicate with the IoT network for attributes collection via a secure channel.      
    
    \item Similar to ABAC-PMF, the AMF also acts as the full Blockchain node to manage the access request and communicate with the Blockchain securely.

    \item Among all the other components, the Blockchain serves as a critical component that stores and manages attributes, ABAC policies, and user requests for access to IoT resources. Additionally, no adversary can control more than 51\% of the network nodes to launch the 51\% attack. Additionally, the policies are secure, and tamper-proof maintained on the Blockchain distributed ledger.

    \item Users are light Blockchain nodes that may impersonate malicious users to get unauthorised access to IoT devices by possessing the users' attributes. Additionally, the attacker can intercept communication between the AMF component and users. However, communication between users and the IoT network is secure after the users obtain the secure access ticket. 
    
\end{itemize}

\section{Proposed Blockchain-based authorisation framework} \label{proposedauthorisation}

Following the network and attacker models, this section presents our proposed flexible and dynamic Blockchain-based authorisation mechanism, which employs the ABAC model to manage users' access requests to IoT network resources based on fine-grained policies in a secure manner. Given that our proposed authorisation process uses the ABAC model to define access control policies for resources, it is equally important to discuss the ABAC model's components, requirements, benefits, and working mechanism as essentials.

\subsection{ABAC model essentials}
ABAC is an authorisation paradigm that determines access by evaluating an individual's attributes (or characteristics) rather than their roles. An ABAC is a logical access control model that emerged from several classic access control approaches such as access control lists (ACLs) and role-based access control (RBAC). ABAC was developed initially to ensure that resources are protected against unauthorised users and behaviours, such as those that do not follow organisational security regulations. 

\begin{figure}[h]
    \centering
    \includegraphics[width=8.5cm, height=3cm]{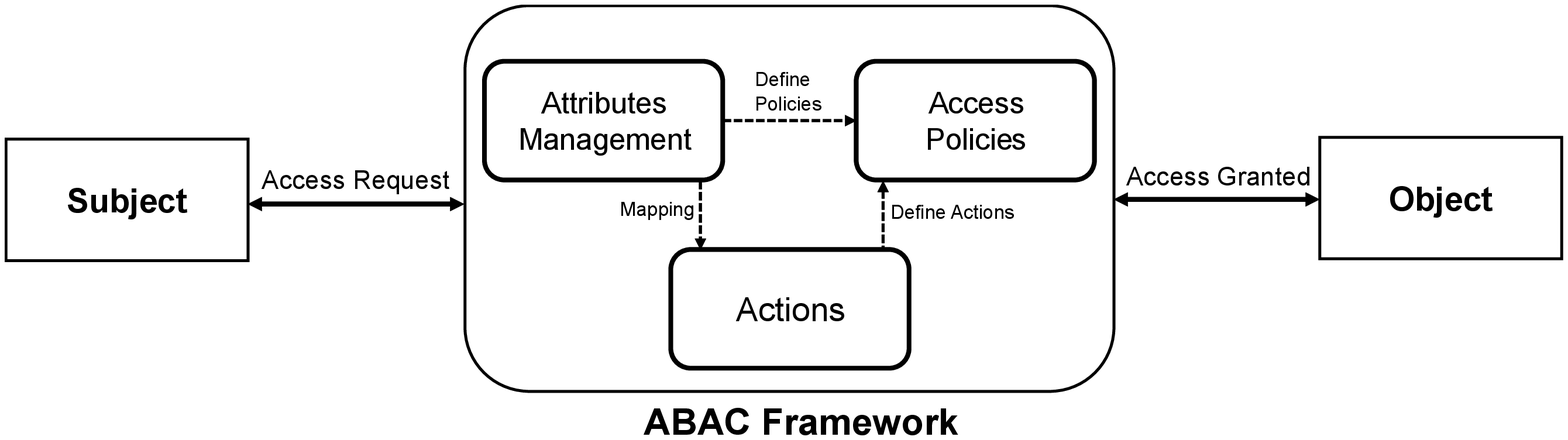}
    \caption{ABAC framework}
    \label{ABACframework}
\end{figure}

There are three primary components in the ABAC model: subject, object, and action, for which an organisation creates access policies to impose access decisions during an access activity. For example, the subject may be a user or an individual who wants to access the system's objects or resources to perform desired actions such as reading or writing. As the name implies, the ABAC model defines access control policies by referencing the attributes of the components involved in an access event. As a result, the primary requirement is to determine how to gather, examine, and define the component's attributes and their interactions in an environment and to build a set of regulations that will define which subjects are allowed to access objects depending on the presence of specified criteria. The working mechanism of an ABAC model is to compare the attributes of the components to stated policies and then determine which attribute mapping from subject to object is authorised to perform an action successfully. Fig. \ref{ABACframework} illustrates the framework of the ABAC model, which consists of related components and their underlying working mechanism.

\subsection{Authorisation process}

Our proposed authorisation mechanism for IoT devices is based on an Ethereum-based Blockchain network, in which access is granted to users who possess specific attributes (subject, object, and environment) defined in access control policies based on the ABAC model. Furthermore, we built access management and ABAC-policy management frameworks based on smart contracts that integrate and communicate with the Ethereum Blockchain in order to manage users access requests and policies, respectively. To better understand how our proposed authorisation process works in its whole, we divide it into the following operational processes and propose algorithms for each one. Table. \ref{tab1} shows the notations and descriptions used in the Algorithms \ref{alg:systemparameters} - \ref{alg:policies}.

\begin{table}[t]
\caption{Notations used in the proposed algorithms}
\begin{center}

\resizebox{\columnwidth}{!}{%
\tiny

\begin{tabular}{|c|c|}

\cline{1-2}
\textbf{Notations} & \textbf{Description}  \\ \cline{1-2}
$\text{\textit{r}}$& Random seed    \\ \cline{1-2}
$\text{\textit{PR}}$& Private key    \\ \cline{1-2}
$\text{\textit{PK}}$& Public key    \\ \cline{1-2}
$\text{\textit{EAddr}}$& Ethereum address    \\ \cline{1-2}
$\text{\textit{BC}}$& Blockchain    \\ \cline{1-2}
$\text{\textit{$U_{i}$} = \{$u_1$, $u_2$, $u_3$, \dots, $u_n$\}}$& Users    \\ \cline{1-2}
$\text{\textit{$D_{i}$} = \{$d_1$, $d_2$, $d_3$, \dots, $d_n$\}}$& IoT devices    \\ \cline{1-2}

$\text{\textit{$M_{Addr}$}}$&  Miner address    \\ \cline{1-2}

$\text{\textit{$M_{ID}$}}$&  Miner ID    \\ \cline{1-2}

$\text{\textit{$D_{Addr}$}}$&  IoT device address    \\ \cline{1-2}

$\text{\textit{$D_{ID}$}}$&  IoT device ID    \\ \cline{1-2}

$\text{\textit{T}}$&  Token    \\ \cline{1-2}

$\text{\textit{$T_{Sig}$}}$&  Signed Token    \\ \cline{1-2}

$\text{\textit{SA}}$& Subject attributes    \\ \cline{1-2}
$\text{\textit{OA}}$& Object attributes    \\ \cline{1-2}
$\text{\textit{EA}}$& Environmental attributes    \\ \cline{1-2}
$\text{\textit{$S_{SA}$}}$& Smart contract for adding subject attributes    \\ \cline{1-2}
$\text{\textit{$S_{OA}$}}$& Smart contract for adding abject attributes    \\ \cline{1-2}
$\text{\textit{$S_{SE}$}}$& Smart contract for adding environmental attributes    \\ \cline{1-2}

$\text{\textit{ID}}$& Unique identifier    \\ \cline{1-2}

$\text{\textit{$a^s_i$}}$&  Individual subject attribute    \\ \cline{1-2}

$\text{\textit{$a^o_i$}}$&  Individual object attribute    \\ \cline{1-2}

$\text{\textit{$a^e_i$}}$&  Individual environment attribute    \\ \cline{1-2}

$\text{\textit{$A^S_i$}}$&  A set of subject attributes    \\ \cline{1-2}

$\text{\textit{$A^O_i$}}$&  A set of object attributes    \\ \cline{1-2}

$\text{\textit{$A^E_i$}}$&  A set of environmental attributes    \\ \cline{1-2}

$\text{\textit{t}}$&  Time \\ \cline{1-2}
$\text{\textit{Loc}}$&  Location \\ \cline{1-2}
$\text{\textit{actv}}$&  Activity \\ \cline{1-2}

$\text{\textit{$T_{x}$}}$&  Blockchain transaction    \\ \cline{1-2}

$\text{\textit{act}}$&  Actions    \\ \cline{1-2}

$\text{\textit{$P_{i}$}}$&  Policies    \\ \cline{1-2}

$\text{\textit{$Tx_{Req}$}}$&  Access request    \\ \cline{1-2}

$\text{\textit{$Att_{i}$}}$&  Attributes set (subject, object, environmental)   \\ \cline{1-2}

$\text{\textit{d}}$&  Decision \\ \cline{1-2}

$\text{\textit{at}}$&  Action taken \\ \cline{1-2}

$\text{\textit{LT}}$&  Lookup table \\ \cline{1-2}

$\text{\textit{$T_{Acc}$}}$&  Access ticket \\ \cline{1-2}

\end{tabular}
}
\label{tab1}
\end{center}
\vspace{-6mm}

\end{table}

\subsubsection{Initialisation of system parameters}

In our proposed mechanism, the authorisation process for IoT devices begins with generating the system parameters for the Ethereum Blockchain, such as Ethereum addresses \textit{EAddrs} and a key pair comprised of a public key \textit{PK} and a private key \textit{PR}. Both \textit{EAddrs} and \textit{PKs} are then uploaded to the Blockchain \textit{BC} for interaction with other system components, such as users and IoT network, as illustrated in the network model. The users are represented as \textit{U} = \{$u_1$, $u_2$, $u_3$, \dots, $u_n$\}, where \textit{U} = \{$u_{i}$ $\vert$ \textit{i} $\in$ \textit{I}\}, for which \textit{I} = \{1, \dots, \textit{n}\}. Further, IoT network consists of number of heterogeneous devices and represented as \textit{D} = \{$d_1$, $d_2$, $d_3$, \dots, $d_n$\}, where \textit{D} = \{$d_{i}$ $\vert$ \textit{i} $\in$ \textit{I}\}, for which \textit{I} = \{1, \dots, \textit{n}\}. Private keys \textit{PRs}, on the other hand, are kept secret by everyone and are not shared with anyone in the network. We send the private keys over a secure channel to both users and IoT devices, assuming that the adversaries do not have access to that channel. To demonstrate how these keys are used in Blockchain-based IoT networks, each IoT device is assigned a unique \textit{PK} and \textit{PR} key pair to generate and validate transaction signatures, respectively. Furthermore, each IoT device and the user has their own \textit{EAs} in the Ethereum Blockchain architecture. The \textit{EAs} are the public addresses, which are the last 20 bytes of the keccak hash of the public key. These parameters are created utilising an off-chain computation process in our proposed mechanism; however, uploading these parameters to the Blockchain requires an on-chain computation process. The entire procedure of initialising system parameters by AA in our system setup is shown in Algorithm \ref{alg:systemparameters}.

\begin{algorithm}[hbt!]
\caption{Initialisation of Systems Parameters}\label{alg:systemparameters}
\textbf{Input:} \textit{r}, \textit{BC}, \textit{$U_{i}$}, \textit{$D_{i}$}  \\
\textbf{Output:} \textit{PKs}, \textit{PRs}, \textit{EAddrs}

\begin{algorithmic}[1]
\Procedure{Initialisation of Systems Parameters} {Random seed}
\State Generate Private Keys, \textit{PRs} $\leftarrow$ SHA256 (\textit{r})
\State Generate Public Keys, \textit{PKs} $\leftarrow$ secp256k1 (\textit{PRs})
\State Generate Ethereum Addresses, \textit{EAddrs} $\leftarrow$ keccak256(\textit{PKs}) 
\State Send \textit{PRs} to the respective users \textit{$U_{i}$} = \{$u_1$, $u_2$, $u_3$, \dots, $u_n$\} and IoT devices \textit{$D_{i}$} = \{$d_1$, $d_2$, $d_3$, \dots, $d_n$\}
\State Upload \textit{PKs} and \textit{EAs} to the Blockchain \textit{BC}
\EndProcedure
\end{algorithmic}
\end{algorithm}

\subsubsection{Define and registration of attributes within an IoT network}

Following the generation and uploading of public parameters to the Blockchain, the next step of the proposed authorisation mechanism is to initialise and register the attributes (i.e. subject, object) utilised in the Blockchain-based IoT network. In general, users and IoT devices must initialise and register attributes before our proposed authorisation, as ABAC-PMF will eventually use these attributes to build policies for IoT device access. For example, to initialise and register subject attributes \textit{SA} and object attributes \textit{OA} on the Blockchain, please refer to our existing Ethereum Blockchain-based decentralised authentication work for IoT networks \citep{hameed2021formally}, in which we used tokens composed of subject and object attributes to authenticate IoT devices. In this work, each IoT device is assigned with a unique signature-based token that is validated by designated miner devices before authentication and registration of those attributes to the Blockchain. 

Additionally, we use a specific piece of information called context-information that is obtained from the deployed IoT environment to initialise and register environmental attributes \textit{EA}. The context information is commonly expressed in the form of semantic data that contains relevant information about the deployed environment and is easily comprehended and interpreted by people. Please refer to our work \citep{hameed2021context} for detailed information on the initialisation, collection, and registration of context information in IoT networks used as environmental attributes.

The Algorithm \ref{alg:attributes} summarises the process of initialising and registering the \textit{SA}, \textit{OA} and \textit{EA} with IoT networks, as described in our previous works.

\begin{algorithm}[hbt!]
\caption{Define and Registration of Attributes}\label{alg:attributes}
\textbf{Input:} \textit{PK}, \textit{PR}, \textit{EAddrs},  \textit{IDs(N)}, \textit{t},  \textit{Loc}, \textit{Actv} \\
\textbf{Output:}  \textit{SA} ,  \textit{OA}, \textit{EA}

\begin{algorithmic}[1]
\Procedure{Define and Registration of Attributes}{}
\State Subject Attributes, $S^a_N$ $\leftarrow$ \{$s^a_1$, $s^a_2$, $s^a_3$, \dots, $s^a_N$\}
     \For{each \texttt{$S^a_N$}}
     
     \If  {\textit{$M_{Addr}$} $\neq$ existing.EAs(\textit{$M_{Addr}$)} \& $M_{ID}$ $\neq$ existing.IDs(\textit{$M_{ID}$)} } 
    \State $s^a_i$ $\leftarrow$ \textit{$M_{Addr}$}, \textit{$M_{ID}$}
    \State Subject Attributes Registration($s^a_i$)
        \Else
        
        $s^a_i$ $\leftarrow$ alreadyRegistered() 
    \EndIf
    \EndFor

\State Object Attributes, $O^a_N$ $\leftarrow$ \{$o^a_1$, $o^a_2$, $o^a_3$, \dots, $o^a_N$\}
\For{each \texttt{$O^a_N$}}
    
    \If  {\textit{$D_{Addr}$} $\neq$ existing.EAs(\textit{$D_{Addr}$}) \& ($D_{ID}$, \textit{$M_{ID}$)}  $\neq$ existing.IDs(\textit{$D_{ID}$,$M_{ID}$)}}
    \State $o^a_i$ $\leftarrow$ \textit{$D_{Addr}$}, \textit{$D_{ID}$}, \textit{$M_{ID}$}
     \State  \textit{T} $\leftarrow$ Token Generation ($o^i_a$)
    \State \textit{$T_{Sig}$} $\leftarrow$ $Signature_{PR} (\textit{T})$
    \State Object Attributes Registration ($o^a_i$)
    \Else
    
    $o^a_i$ $\leftarrow$ alreadyRegistered() 
    \EndIf
\EndFor

\State Environmental Attributes, $E^a_N$ $\leftarrow$ \{$e^a_1$, $e^a_2$, $e^a_3$, \dots, $e^a_N$\}
\For{each \texttt{$E^a_N$}}
        \If  {\textit{$D_{ID}$ $\neq$ existing.IDs(\textit{$D_{ID}$)}} \& \textit{t} $\neq$ Current.time() \& \textit{Loc} $\neq$ Location(Subject) \& \textit{Actv} $\neq$ Activity.List(Object)} 

        \State $e^a_i$ $\leftarrow$ \textit{$D_{ID}$}, \textit{t}, \textit{Loc},\textit{Actv}
       \State Environment Attributes Registration ($e^a_i$)
\Else 

\State $o^a_i$ $\leftarrow$ alreadyRegistered()
        \EndIf
        
\EndFor
\EndProcedure
\end{algorithmic}
\end{algorithm}

\subsubsection{Registration and management of attributes in Blockchain network}\label{regisandmanageattributes}

After defining and registering the \textit{SA}, \textit{OA} and \textit{EA} with the IoT network, the next step in our proposed authorisation mechanism is to register and manage the attributes on Ethereum Blockchain.  As described earlier in the network model, we propose an ABAC-PMF framework with dual responsibilities. First, it is responsible for collecting, storing, and managing the attributes in the Blockchain that are collected via the AMA component from the IoT network. Second, using the PMA component is responsible for defining, storing, and managing access policies in the Blockchain.

To begin, the AMA component of the ABAC-PMF framework interacts with the IoT network to collect the attributes recorded on the Blockchain. It is important to note that this Blockchain is separate from our primary Ethereum Blockchain, which the IoT network uses to define and register attributes for IoT device authentication. The AMA component comprises three types of smart contracts: $S_{SA}$, $S_{OA}$, $S_{EA}$, which are used to collect, store, and manage attributes relating to the subject, object, and environment, respectively. The following are the details of the smarts contracts, including the type of attributes, functions and their working mechanism:

\begin{itemize}
    \item \textbf{Subject-related attributes:} We create a smart contract called $S_{SA}$ to collect, store and manage subject-related attributes on the Blockchain. The subjects are often the individuals or users who wish to access system resources or objects, followed by the policies specifying the types of actions they are permitted to execute, such as read or write. Each subject in the network is distinguished from others by a unique identifier, such as an \textit{ID}. However, one \textit{ID} may contain several attributes to demonstrate the possession of attributes associated with the same \textit{ID} through the use of different access policies. For instance, an administrator may have multiple attributes that enable him or her to access the organisation's different resources for controlling the database and network operations. 
    
    We define the subject's attributes as the \textit{SID}, Ethereum Address (\textit{EAddr}), \textit{Name}, \textit{Role} and \textit{Location} in our scenario. Each attribute is a key-value pair with two values, such as the name and value of the attribute, and is denoted by $a^s_i$ = $<name, value>$. A collection of attributes associated with a given subject is denoted by $A^S_i$ = \{$a^s_1$, $a^s_2$, $a^s_3$, \dots, $a^s_N$\}. For instance, the subject \textit{k} has the attributes that are denoted by $A^s_k$ = \{\textit{SID}, \textit{EA}, \textit{Name}, \textit{Role} and \textit{Location}\}.   
    
    The smart contract $S_{SA}$ is deployed exclusively by the administrators of subjects. In this case, the AMA is responsible for executing the $S_{SA}$ on the Blockchain.
    
    \item \textbf{Object-related attributes:} Similar to $S_{OA}$, we create a smart contract called $S_{OA}$ to collect, store and manage object-related attributes on the Blockchain. The objects are the system resources to which the subject wants to gain access, including files, services or applications, application programming interfaces (APIs), or hardware resources. In our case, we focused specifically on IoT devices as the objects that need to be protected from attackers via unauthorised access. Similarly to subjects, each object is distinct due to its unique identity, such as the IP address of IoT devices. Additionally, each object can have several attributes associated with its ID, indicating its possession of those attributes. We refer to the object \textit{ID} as \textit{OID} for simplicity. 
    
    We describe the object attributes as \textit{OID}, Ethereum Address (\textit{EAddr}), \textit{Object name} and \textit{object type} in our case.  A collection of attributes associated with a given object is denoted by $A^O_i$ = \{$a^o_1$, $a^o_2$, $a^o_3$, \dots, $a^o_N$\}.  For instance, the object \textit{k} has the attributes that are denoted by $A^o_k$ = \{\textit{OID}, \textit{EA}, \textit{Name}, \textit{Object name} and \textit{Object type}\}.
    
    The smart contract $S_{OA}$ is deployed exclusively by object owners, and in our case, the AMA is responsible for executing the $S_{OA}$ on the Blockchain.

    \item \textbf{Environmental-related attributes:} To establish dynamic and flexible access rights for system resources, environmental factors are critical in the broader context of each access request. Environmental attributes, more precisely, are contextual data that provide helpful information about environmental elements such as subjects and objects in terms of their behaviour pattern and interaction with the deployed environment. For example, contextual information may include the following: time, location (for both subject and object), object behaviour pattern, critical operations requirements, authentication strength, encryption strength, and so forth. In our case, we specify the time, subject location, object behaviour pattern, and authentication status as environmental attributes for the purpose of achieving dynamic access rights to objects. The following is a description of the environmental attributes used in our case: (i) \textit{Time} is defined as the time allotted to each subject to conduct the operations indicated in the policy on the object, (ii) \textit{Subject location} refers to a particular location from which resources can be accessed, (iii) \textit{Object behaviour pattern} defines the object's current state, such as an IoT device that has been compromised and has become malfunctioning, (iv) Finally, an \textit{authentication status} specifies the information or state of an IoT device's authenticity or non-authenticity.
    
    A collection of attributes associated with a deployed environments is denoted by $A^E_i$ = \{$a^e_1$, $a^e_2$, $a^e_3$, \dots, $a^e_N$\}. For example, a deployed environment \textit{k} has the attributes $A^e_k$ = \{\textit{Time}, \textit{Subject location}, \textit{Object behaviour}, \textit{Authentication status}\} for both subject and object.

    Other environment-related attributes, such as critical operations or activities (e.g., data sense, calculation, transmission, and temperature control), are included in access policies as core actions. Similarly to the deployment of $S_{SA}$ and $S_{OA}$ smart contracts, the AMA executes and deploys the $S_{EA}$ smart contract on the Blockchain.

    \item \textbf{Management of attributes:} Apart from registering attributes on the Blockchain network via the AMA component, we defined and implemented smart contract functionalities for updating and revoking subject, object, and environment-related attributes.

\end{itemize}

Table \ref{tableexamples} presents an example of the subject, object, and environmental attributes in key-value forms used by our proposed authorisation scheme.

\begin{table}[h]
\centering
\caption{Example of attributes utilised in our proposed scheme}
\resizebox{\columnwidth}{!}{%
\begin{tabular}{|l|l|l|}
\hline
 \textbf{\makecell{Subject \\ Attributes}}& \textbf{\makecell{Object \\ Attributes}}  & \textbf{\makecell{Environmental \\ Attributes}}  \\ \hline
\textit{$<SID: 321>$}  & \textit{$<OID: 325>$}  & \textit{$<Time:E_{Time}-S_{Time}>$} \\ \hline
\textit{$<EAddr: 0\times43a6cb>$} & \textit{$<EAddr: 0\times6d34e2>$} & \textit{$<Sub.location:East.AUS>$}   \\ \hline
\textit{$<Name: Charlie >$} & \textit{$<Obj.Name: Lock >$} & \textit{$<Obj.behaviour:Malicious>$} \\ \hline
\textit{$<Role: User>$} & \textit{$<Obj.Type: Security>$} & \textit{$<Auth.status:Auth/Non.Auth>$} \\ \hline
\textit{$<Location: East.AUS>$} &&\\ \hline
\end{tabular}}
\label{tableexamples}
\end{table}

\subsubsection{Define and management of access control policies in Blockchain network}

\begin{algorithm}[hbt!]
\caption{Define and Management of Policies}\label{alg:policies}
\textbf{Input:} $A^S_i$ = \{$a^s_1$, $a^s_2$, $a^s_3$, \dots, $a^s_N$\}, $A^O_i$ = \{$a^s_1$, $a^s_2$, $a^s_3$, \dots, $a^s_N$\}, $A^E_i$ = \{$a^s_1$, $a^s_2$, $a^s_3$, \dots, $a^s_N$\}, \textit{act} = \{\textit{Read}, \textit{Write}, \textit{Execute}\} \\
\textbf{Output:} $P_{i}$ = \{$P_{1}$, $P_{2}$, $P_{3}$, \dots, $P_{n}$\}

\begin{algorithmic}[1]

\Procedure{addPolicy()}{}
\State Initialisation of Policies $P_{i}$
\State Collection of Attributes \{$A^S_i$, $A^O_i$,$A^E_i$\} from \textit{AMA}
\State Define Actions $act_{i}$
 \For{each $P_{i}$}
\State Initialise $Tx_{P_{i}}$ = \{$A^S_i$,$A^O_i$,$A^E_i$,$act_{i}$\}
\State Store $P_{i}$ on Blockchain
 \EndFor
\EndProcedure

\Procedure{UpdatePolicy()}{}
\If{$P_{i}$ $\leftarrow$ searchPolicy($ID_{i}$)}
\State Update Actions $act_{i}$
\State $P_{i}$ $\leftarrow$ addPolicy()
\State Store $P_{i}$ on Blockchain
\Else
\State  $P_{i}$ $\leftarrow$ notExisted()
\EndIf
\EndProcedure

\Procedure{RevokePolicy()}{}
\If{$P_{i}$ $\leftarrow$  searchPolicy($ID_{i}$)}
\State Revoke Policy $P_{i}$
\Else
\State  $P_{i}$ $\leftarrow$ notExisted()
\EndIf
\EndProcedure

\end{algorithmic}
\end{algorithm}

After successfully registering attributes, the next step in the proposed authorisation scheme is defining and storing the access control policies on the Blockchain. As we mentioned earlier, the ABAC-PMF has a second important component named PMA, responsible for defining, storing and managing the policies on the Blockchain. In our proposed scheme, the PMA component defines the more fine-grained policies by adding more attributes related to subject, object and environment to the Blockchain, thereby achieving more flexible and dynamic access control to the critical resources of the IoT network. The PMA component includes the following functions, all of which are built-in smart contracts, for defining and managing the policies on the Blockchain. The algorithm \ref{alg:policies} demonstrates how access control policies are defined and managed on the Blockchain. 

\begin{itemize}

\item \textbf{Adding the new policies:} To add a new policy to the Blockchain, the PMA first interacts with the AMA to get the attributes (subject, object, and environment) and then initiates a Blockchain transaction ${T_{x}}$ containing the attributes $A^a_k$ = \{$A^s_k$, $A^o_k$, $A^e_k$\} and actions, \textit{act} = \{\textit{Read}, \textit{Write}, \textit{Execute}\} to be defined for giving specific access to IoT resources. Each policy is stored to the Blockchain with the unique ID, called $P_{ID}$, which helps to identify the policies uniquely and to retrieve the policies from the Blockchain. Moreover, we evaluate only those actions that are relevant to user access to IoT resources, such as reading, writing, and executing data using those resources, when defining policies. However, network administrator tasks such as updating device configuration, creating and deleting devices from registries, and controlling the network are not covered in policy definition.

Furthermore, to make policies efficient, simple, small in size, and fast to process and retrieve, we create a structure for policies that is distinct from the \citep{putra2021trust} scheme, as this scheme combined all access control parameters and actions taken into a single policy definition, impeding the policy structure.

\item \textbf{Updating the existing policies:} As with adding new policies to the Blockchain, the PMA can also update existing policies in the ABAC-PMF. However, policy updates are only required when an authorised user requires additional actions beyond those already specified in the policy. For example, if a user is initially assigned only read rights but later requires special privileges such as writing, an updated policy is required. The PMA performs a simple search operation utilising the unique ID issued to each policy to update the policy. Although alternative search methods are employed in the current system, searching policies using those methods takes longer and consumes more gas \citep{zhang2020attribute}.

\item \textbf{Revoking the existing policies:} Along with adding and updating the policies to the Blockchain, the PMA also has the functionality to revoke the existing policies from the Blockchain network. However, policy revocation is required only when some malicious behaviour is detected, and the PMA component gets a notification from the AMA component to take appropriate action. The policy revocation may be necessary for two situations: if the user has become malicious or if specific devices have been compromised.

\end{itemize}

Table \ref{tab:policies} illustrating several policies for the ABAC model. Each policy is defined to grant users access to IoT devices based on defined single or multiple attributes and related device actions. For instance, policy $P_{1}$ is specified to provide access to users based on a single set of attributes, such as subject \textit{ID} $<SID:123>$, object \textit{ID} $<OID:112>$, subject location $<Sub.location:West.AUS>$.  Furthermore, numerous attributes with distinct sets of actions may be used to provide users with more fine-grained access to IoT devices. For example,  policy $P_{2}$ is specified for giving the access to users on the basis of multiple attributes such as subject \textit{ID} $<SID:123>$ and \textit{Object Name}  $<Name:Charlie>$ to access the object with object \textit{ID} $<OID:112>$ and object name $<Obj.Name:Thermostat>$ and environmental attributes subject location $<Sub.location:West.AUS>$ and object behaviour $<NonMalicious>$ with defined actions such as \textit{Read} and \textit{Write}.

\begin{table}[!htp]
\centering
\caption{An example of ABAC model policies with defined attributes and actions}
\resizebox{\columnwidth}{!}{%
\begin{tabular}{|c|c|c|c|c|}
\hline
\multirow{2}{*}{\textbf{Policies}} & \multicolumn{3}{c|}{\textbf{Attributes}} & \multirow{2}{*}{\textbf{Actions}} \\ \cline{2-4}
&      \textbf{Subject}    &     \textbf{Object}     & \textbf{Environment}          &                            \\ \hline

                                        $P_{1}$              &   $<\textit{SID:123}>$       &  $<\textit{OID:112}>$        &       $<\textit{Sub.location: West.AUS}>$   &   Read                         \\ \hline

                       $P_{2}$     & \thead{$<\textit{SID:123}>$ \\ $<\textit{Name:Charlie}>$}          &
                       \thead{$<\textit{OID:112}>$ \\ $<Obj.Name: Thermostat:>$  }
                       
                               &   
                               
                     \thead{$<\textit{Sub.location: West.AUS}>$ \\
                    $<Obj.Behaviour: Non Malicious>$ }          
                               
                                    &   \thead{Read \\Execute}                        \\ \hline

                          $P_{3}$  & 
                          
                          \thead{$<EAddr:0\times5fe321>$ \\ $<Role: Admin>$ }
                          
                               & \thead{$<\textit{OID:167}>$ \\ $<EAddr:0\times12a34f>$ \\ $<Obj.Name: Thermostat:>$}          &

                               \thead{ $<Obj.Behaviour: Non Malicious>$ \\ $<Auth.Status:Auth>$ }
                               
                                      & \thead{Read \\ Write \\ Execute}

                          \\ \hline
                          
                          $P_{4}$  & $<EAddr:0\times5fe321>$         & $<OID:345>$         &  $<Time:E_{Time}-S_{Time}>$        & Read                           \\ \hline

                          $P_{5}$  & $<\textit{SID:145}>$         & $<Obj.Type: Security>$         &  $<Sub.location:East.AUS>$        & Write                            \\ \hline
\end{tabular}}
\label{tab:policies}
\end{table}

\subsubsection{Validation of access request}

After successfully registering attributes and managing policies on the Blockchain through ABAC-PMF, an important and last step in the proposed authorisation scheme is to grant users access to the resources of IoT networks that meet the access policy's specified attributes. As previously stated, we designed a separate framework called AMF that makes use of smart contract functions to perform a variety of tasks, including evaluating and managing user requests, retrieving relevant policies, allocating resources, taking actions on user requests, making decisions, and auditing user requests and resources. The Algorithm depicts the entire process of evaluating user requests to assign resources based on stated policies with specific Boolean attribute rule sets in the access policy. 

To begin the process of authorising users to use the resources of an IoT network, a user must first submit an access request to AMF via the request transaction $Tx_{Req}$.  A $Tx_{Req}$ is a signed transaction that contains the users' attributes (i.e., subject attributes), the access resource identifier $O_{ID}$, and the requested action \textit{act} to perform on the accessible resources. To sign the transaction $Tx_{Req}$, the user used his or her private key \textit{PR}, which is generated during the initialisation step of system parameters. It is important to note that before a user can access the particular IoT resources, the subject attributes must be registered and stored on the Blockchain. After receiving the $Tx_{Req}$ from the users, AMF communicates with the Blockchain to validate the $Tx_{Req}$ using the saved public key \textit{PK} and determines whether or not necessary attributes have been registered and stored to the Blockchain. After successfully validating $Tx_{Req}$ in terms of subject attributes via Blockchain, the AMF framework retrieves the relevant policy $P_{i}$ for which the access request was made and maintains a lookup table containing the user identification $SID_{i}$, the policy identification $P_{ID}$, the decision made on the request \textit{d} = \{\textit{Approved, \textit{Denied}}\}, and further setting the restrictions as action taken \textit{at} based on the user's behaviour. After successfully storing and maintaining the required information in the lookup table, the AMF issues an access ticket $T_{Acc}$ containing the lookup table information and issues it to users requesting access to the IoT network's requested resources. The $T_{Acc}$ is denoted by the \{$SID_{i}$, $P_{ID}$, \textit{d}, \textit{at}\}. Finally, after obtaining the $T_{Acc}$ from the AMF, users communicate with the IoT network via a secure communication channel by presenting the issued token to the IoT network in order to gain access to the specific IoT device and data. The communication route between users and the IoT network is assumed to be secure and maintained by the AA in this case.

\begin{algorithm}[hbt!]
\caption{Validation of Access Request}\label{alg:policies}
\textbf{Input:}  $Tx_{Req}$, \textit{$Att_{i}$} = \{$A^S_i$, $A^O_i$, $A^E_i$\}, $P_{i}$, \textit{d}, \textit{at}, \textit{LT}\\
\textbf{Output:} $T_{Acc}$

\begin{algorithmic}[1]

\Procedure{Validation of Access Request}{}

\State Generate Access Request $Tx_{Req}$ $\leftarrow$ \textit{$Sig_{PR}$($A^S_i$)}
\State Validate Access Request \textit{Val} $\leftarrow$ \textit{$Validate_{PK}(Tx_{Req})$}
\If{\textit{Val} == \textit{Approved} \textbf{and} \textit{at} == $\emptyset$ }
\State Fetch Policies $P_{i}$
\State Fetch Attributes \textit{$Att_{i}$}
\State Issue $T_{Acc}$ $\leftarrow$ \{$S_{ID_i}$, $P_{ID}$, \textit{d}, \textit{at}\}
\State Update \textit{LT}
\Else
\State $A^S_i$ $\leftarrow$ NotExisted()
\State \textit{at} [$A^S_i$] $\leftarrow$ Blocked
\State Update \textit{LT}
\EndIf

\EndProcedure

\end{algorithmic}
\end{algorithm}

The main aim of designing the AMF framework in the authorisation scheme through the use of Blockchain's smart contract features is twofold: (i) First, to manage each user request for access to IoT resources separately and securely, thereby preserving the privacy of their information and data; and (ii) Second, to make the proposed scheme auditable in terms of user requests for access to IoT resources, access control policies, and attributes, as well as to further determine user behaviour for access decision enforcement.

\begin{figure*}[t]
    \centering
    \includegraphics[scale=0.52]{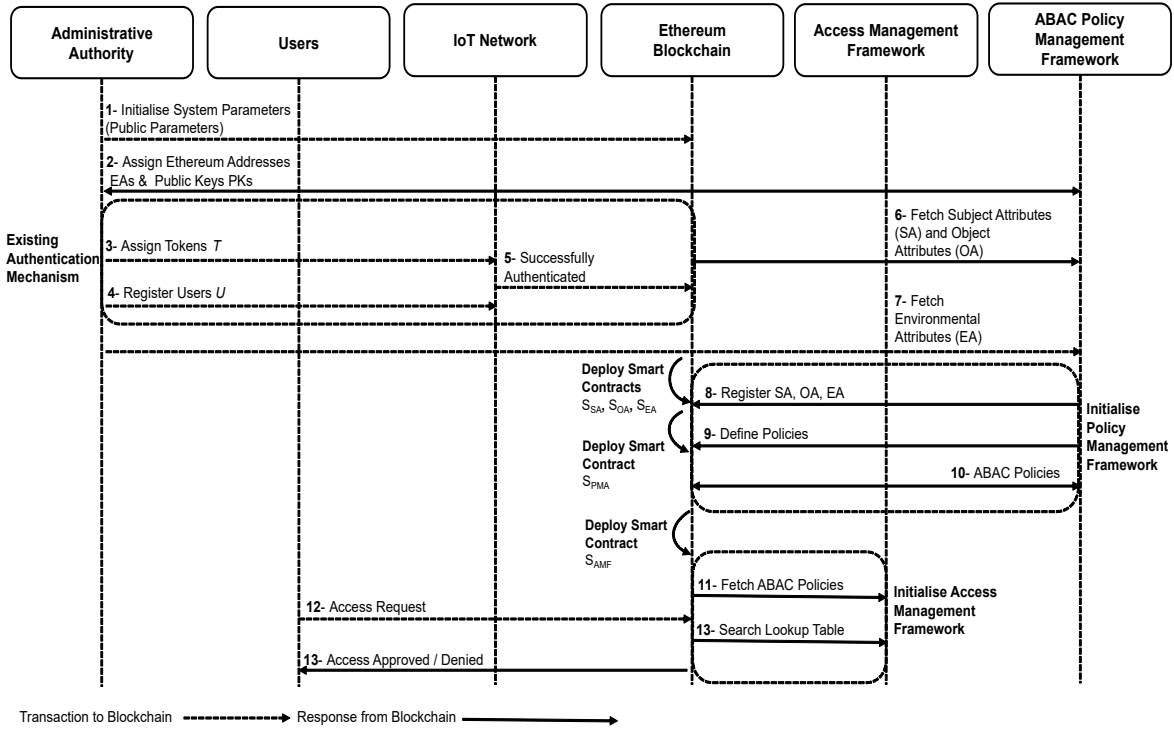}
    \caption{Overall Execution Flow of Proposed Blockchain-based Authorisation Mechanism }
    \label{executionflow}
\end{figure*}

\subsection{Execution flow}

In a decentralised environment, the authorisation procedure for users accessing IoT devices on the Blockchain network is performed using various interconnected modules that provide timely responses and act accordingly. Our proposed Blockchain-based secure, flexible and dynamic authorisation mechanism for IoT devices consists of the following critical interacting components: administrative authority, users, IoT network, Ethereum Blockchain, access management framework and ABAC-policy management framework. These components enable authorised users to secure access to IoT devices via the ABAC model built on Ethereum Blockchain. Fig. \ref{executionflow} shows the complete execution flow of the authorisation process for users accessing IoT devices, from attribute registration to policy definition and after that managing secure user access via smart contract features.

The procedure starts with the administrative authority responsible for initialising and managing the overall system parameters such as Ethereum addresses and public keys. After initialisation, the parameters are assigned to the respective authorities, such as AMF and ABAC-PMF, to communicate with the Ethereum Blockchain and other network model components. Furthermore, the administrative authority is also responsible for initialising the IoT network setup, which consists of many heterogeneous IoT devices authenticated via the Ethereum Blockchain.

The next step in the proposed authorisation scheme is to authenticate individuals and IoT devices using the token mechanism and then define and initialise their attributes for use in policy definition. Since our scheme, compared to previous works \citep{zhang2020attribute} \citep{putra2021trust}, uses environment-related attributes to dynamically grant access to IoT devices depending on their behaviour in IoT networks, the administrative authority also stores environment-related attributes in the ABAC-PMF.
  
Following attributes initialisation, the next step is registering these attributes with the ABAC-PMF via smart contract capabilities. Each smart contract is responsible for registering, storing, and managing attributes on the Ethereum Blockchain through the use of the functionality provided in it. After registering attributes in ABAC-PMF, the next step is to define policies based on the registered attributes and prescribed actions for each resource's access level and nature. Similar to attributes, the ABAC-PMF are also handled by smart contracts written to register, store, and manage policies on the Blockchain.

After successfully defining attributes and policies in the ABAC-PMF, the final step is to secure user access to IoT devices via the AMF. The AMF is composed of a lookup table that maps attributes to rules and actions and then determines whether or not users are permitted to access the particular resources to which access is being granted. Additionally, this component ensures the audibility of the user's access to IoT resources in order to ascertain their behaviour and actions on them.

\subsection{Qualitative security analysis}

We performed a qualitative security analysis of the proposed authorisation scheme for IoT networks by considering the achieved security requirements. 

For instance, the benefit of incorporating Blockchain technology into our proposed scheme is that it enables us to achieve decentralisation in order to address the single-point-of-failure issues inherent in conventional authorisation mechanisms. The single-point-of-failure problem arises as a result of compromised system entities such as IoT devices and storage databases, rendering authentication information and access control policies unreachable in a single location. On the contrary, by utilising Blockchain, each distributed party can keep their own database of access control policies and have their own copy of the data, which results in a high level of data availability. Furthermore, the use of distributed ledger technology (DLT) in Blockchain, which is shared and synchronised across several nodes, ensures that access control policies on the ledger are secure and precisely stored through the use of cryptography. In our proposed scheme, once access control policies are stored on Blockchain with the approval of PMA component and network consensus protocols, these constitute an immutable database, making unauthorised policy changes nearly impossible. To ensure the security of the IoT network in terms of protecting IoT devices and their data, we utilise our existing Blockchain-based decentralised authentication scheme to ensure that only legitimate users in the IoT network are permitted to register IoT devices on the Blockchain using authentication tokens (e.g., subject and object attributes). Upon validating authentication tokens by the miner nodes, the IoT devices are registered on the Blockchain network. Besides that, since we used a separate framework (i.e., AMF) in conjunction with Ethereum Blockchain features to manage the requests of each user accessing IoT devices, access tickets are generated and assigned to users for each distinct request and managed appropriately, ensuring the privacy of users' information. Finally, we ensured the auditability of user requests to access IoT devices and the policies associated with them in order to ensure the evaluation of access control policies is auditable. The lookup table implemented in AMF ensures the auditability feature of Blockchain technology which is derived from its immutability and transparency properties.

\section{Implementation and evaluation framework}\label{evaluation}

\begin{figure}[h]
    \centering
    \includegraphics[width=9cm, height=6cm]{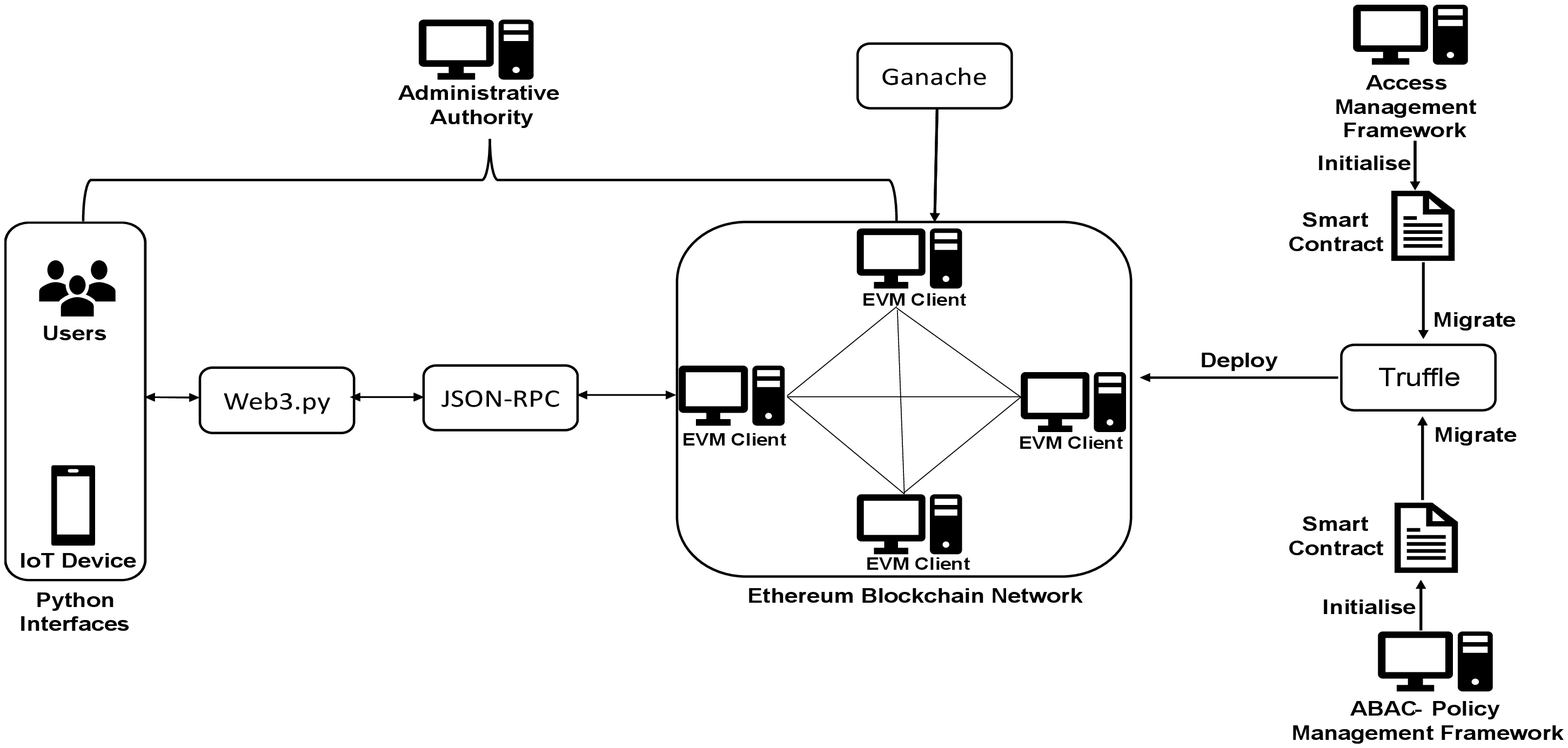}
    \caption{Evaluation framework}
    \label{prototype}
\end{figure}

As illustrated in Fig. \ref{prototype}, our implementation and evaluation framework is built on various technology paradigms, including the Ethereum Blockchain, Python, Web3.py, JSON-RPC, Ganache, Solidity, and Truffle Suite. These technological paradigms assist us in developing and efficiently evaluating the core functions of our proposed authorisation scheme. For instance, we implemented our proposed decentralised prototype for our authorisation scheme on the Ethereum Blockchain because it is a widely used platform for developing decentralised applications (dApps) and further allowing the creation of a secure way to conduct transactions via the elliptic curves cryptography protocol.

The following are the details concerning these paradigms. The Ethereum Blockchain is important to our proposed authorisation method since it enables the efficient deployment and execution of smart contract-defined operations. We developed our prototype interfaces in Python, which is a dynamic and scalable language that runs on a wide variety of platforms. We used the Python-based Web3.py framework, which allowed us to communicate with Ethereum clients and execute functions written in smart contracts. The JavaScript-based JSON-RPC protocol is primarily for client-side application communication. We deployed and tested our proposed authorisation scheme using Ganache, a Blockchain emulator, without setting up the main Ethereum network. Each node acts as a personal Ethereum client. We used solidity programming to create the Ethereum-based smart contracts outlined in our proposed system. Finally, the Truffle Suite is used to create, implement, and execute smart contracts on the Ethereum Blockchain.

\section{Performance analysis and results discussion}\label{performance}

Our experiments are conducted on a system that meets the requirements listed in Table \ref{tab2}. We used a Dell Laptop (Intel Core i7-8650U) running Ubuntu 18.4 64-bit. The 64-bit operating system employed the $\text{{x86\_64}}$ CPU architecture, which has a clock cycle of 2.11 GHz. In addition, our experimental specification contained 16 GB of RAM.

\begin{table}[!h]
\centering
\caption{System specifications}
\begin{adjustbox}{width=0.47\textwidth,center}
\begin{tabular}{|c|c|c|c|c|c|}
\hline

\vtop{\hbox{\strut \textbf{Operating} }\hbox{\strut \textbf{systems}}} & 

\vtop{\hbox{\strut \textbf{Operating systems} }\hbox{\strut \textbf{Mode}}} 

& \textbf{CPU architecture}

& \textbf{CPU clock cycle} & \textbf{RAM}  \\ \hline
       Ubuntu 18.04   &   64-bits        &     $\text{{x86\_64}}$     &       2.11 GHz    &    8 GB               \\ \hline
\end{tabular}
\end{adjustbox}
\label{tab2}
\end{table}

\subsection{Case study: An IoT-based smart home} \label{casestudy}

The rate at which computing elements, particularly in smart homes, are adopted depends on the security level given by the specifically designed applications. An IoT is a significant enabler in the smart home environment, enabling home automation and improving quality of life.  In IoT-based smart homes, smart gadgets, also known as IoT devices, are deployed throughout the house to provide the homeowner with a secure environment and control everything remotely by using the Internet.  For example, Intelligent security alarms and motion detectors can operate intelligently and notify the homeowner if there has been a security breach. In a smart home environment, a house owner sets up and deploys the smart devices with the appropriate access control policies to permit authorised users to act accordingly. However, an adversary always tries to breach the security policies defined by the homeowner to unauthorised access to deployed home devices to get the user's personal information or data. As a result, developing a secure authorisation framework for IoT-enabled smart homes that restricts access to only authorised users becomes increasingly essential.

Given the growing popularity of IoT-based smart home applications for complete house automation, along with achieving the security of IoT devices and data, we illustrate the applicability of our proposed Blockchain-based trustworthy, flexible, fine-grained, and dynamic decentralised authorisation mechanism in an IoT-based smart home scenario. Furthermore, it is important to note that the applicability of our proposed scheme is not limited to the smart home scenario but can be extended to other IoT-based applications such as smart health, smart city, smart transportation, and so on by properly defining the attributes and policies of the relevant applications.

In an IoT-based smart home scenario, secure permission is granted to the user to access the IoT devices based on attribute-based access control policies defined in the Ethereum-based Blockchain. The users act as the light nodes in the proposed Blockchain-based system who want to access IoT devices to perform actions in the IoT network. In comparison, the ABAC-PMF and AMF are critical components of our proposed scheme, as both served as full nodes (e.g., miners), determining user requests and communicating with the Blockchain for the purpose of accessing and granting access control policies.

\subsection{Cost evaluation}

We estimate the cost evaluation in terms of financial costs involved with deploying our proposed scheme on the Ethereum Blockchain and then utilising its functions to provide users with secure and dynamic access control. There are two types of cost used to evaluate any system deployed on the Ethereum Blockchain: deployment cost and execution cost. We computed both types of costs for our proposed authorisation schemes, which are specified in the subsections \ref{deploycost} and \ref{executioncost}.

Furthermore, a user interacting with the Ethereum Blockchain must pay a fee in order to deploy and execute smart contracts using ABIs. In the Ethereum Blockchain, a unit called Gas is used to measure the amount of money required to perform the required functions. The miners set the gas price during the transaction, which is expressed in Gwei. Gas prices (e.g., units) are calculated to execute smart contracts and their associated functions. In the Ethereum Blockchain, gas units are transformed to ether, referred to as the fuel of Blockchain. The more complicated the smart contract function, the more money (e.g., Gas) the user must pay. 

\subsubsection{Deployment cost}\label{deploycost}

The \textit{deployment cost} is referred to as the cost of sending (e.g., deploying) a smart contract to the Ethereum Blockchain. This cost is proportional to the size of the smart contract, which contains a variety of computing functions designed to accomplish a variety of activities. For instance, if there are more functions with high computing demands, such as power and polynomial functions, the smart contract will be enormous in size and have high transaction costs.

We determine the deployment cost of our proposed scheme, which is the cost associated with deploying the various smart contracts on the Ethereum Blockchain. Since our proposed scheme is composed of two primary components, the ABAC and AMF, each of which has a distinct set of smart contracts for performing the authorisation process. For example, the ABAC-PMF component comprises four smart contracts named  $S_{SA}$, $S_{OA}$, $S_{EA}$ and $S_{PMA}$ that manage the attributes and policies on the Blockchain. In contrast, the AMF component consists of only one smart contract named $S_{AMF}$ that manages user requests and the authorisation process. Furthermore, each smart contract is deployed separately via a single transaction; the deployment cost is calculated as the amount of Gas spent by the transaction when it is sent to the Ethereum Blockchain.

\begin{figure}[h]
    \centering
    \includegraphics[width=8cm, height=5cm]{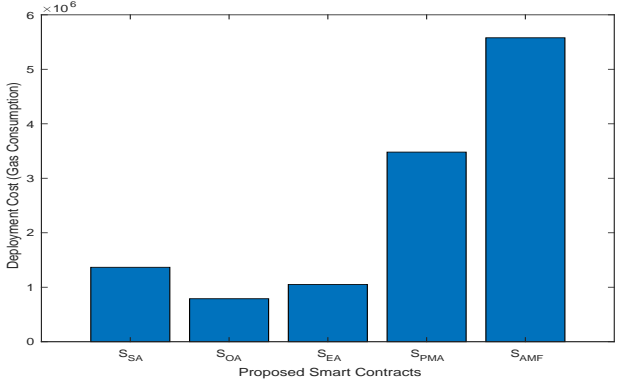}
    \caption{Deployment cost of proposed scheme}
    \label{deploymentcost}
\end{figure}

Fig. \ref{deploymentcost} depicting the deployment cost of our proposed authorisation scheme on the Ethereum Blockchain in terms of gas usage. As can be seen, the cost of deploying the \textit{AMF} component on Blockchain is relatively high because it contains numerous smart contract functions for managing various tasks such as managing user requests, auditability of user requests and policies, managing the dynamic access process, and making further decisions. Additionally, the deployment cost of the \textit{PMA} component on the Ethereum Blockchain is less than that of the \textit{AMF} component, as the \textit{PMA} component is solely responsible for establishing, storing, and managing policies on the Blockchain. Finally, the deployment cost of the \textit{AMA} component in terms of attributes registration (i.e., subject, object, and environment) is lower than that of other components. 

However, we believe that the cost of deploying a system on Blockchain is totally dependent on its complexity in terms of the number of defined functions, their size, and the logic and activities that underpin them. Therefore, we aimed to minimise the deployment cost in our proposed scheme by utilising the finest expertise available for designing code with the fewest possible operations while maintaining the system's functionality.


\subsubsection{Execution cost} \label{executioncost}

The \textit{execution cost} is determined by the cost of the executing computational operations as a result of the transaction. The transaction cost is calculated in terms of deploying the smart contract and the data associated with the transaction.

We calculated the execution cost of performing different operations written in smart contracts in order to realise the process of implementing a secure, dynamic, and flexible authorisation mechanism for IoT devices. We compare the execution cost of our proposed scheme to the existing state-of-the-art Blockchain-based ABAC control schemes \citep{zhang2018smart} \citep{yutaka2019using}. It is important to note that, in contrast to existing schemes\citep{zhang2018smart} \citep{yutaka2019using}, we additionally used environment-related attributes to define dynamic and fine-grained access control policies.

\begin{figure}[h]
    \centering
    \includegraphics[width=8cm, height=5cm]{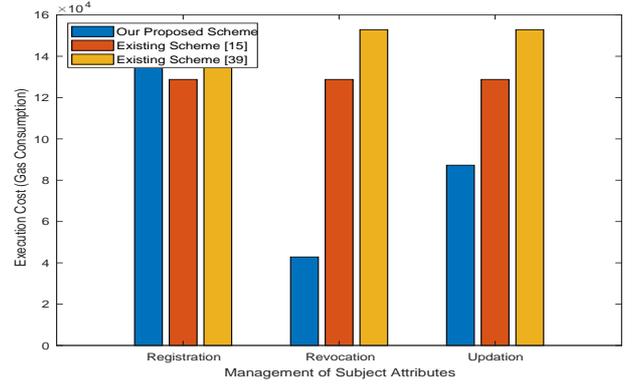}
    \caption{Execution cost for managing subject attributes}
    \label{executioncostsubjectattributes}
\end{figure}

Fig. \ref{executioncostsubjectattributes} depicts the execution cost, in terms of gas consumption, associated with managing subject attributes via the use of several defined functions in smart contract $S_{SA}$. The execution cost of registering the subject attributes is substantially higher in our proposed scheme, as the transaction involves several attributes to be stored on the Blockchain. Moreover, the execution cost of updating the subject attributes is less than the execution cost of revoking the attributes, as revoking requires the specific attributes to be disabled on the Blockchain. Additionally, the cost of executing the managing operations on subject attributes under the scheme \citep{zhang2018smart} is greater than the cost of executing the same operations on subject attributes under the existing system \citep{yutaka2019using}.

\begin{figure}[!ht]
    \centering
    \includegraphics[width=8cm, height=5cm]{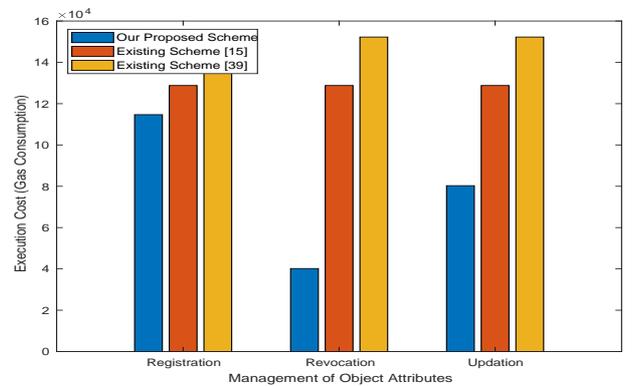}
    \caption{Execution cost for managing object attributes}
    \label{executioncostobjectattributes}
\end{figure}

Fig. \ref{executioncostobjectattributes} depicts the execution cost, in terms of gas consumption, of managing object attributes via the usage of several described functions in a smart contract $S_{OA}$ operated on the Blockchain. Similar to subject attribute registration, the operation of adding object attributes to the Blockchain consumes more Gas in terms of execution cost than other activities. Further, the cost of performing update operations on object properties exceeds the cost of doing revocation operations. Finally, the overall execution cost of operations on object attributes is lower for the existing scheme \citep{zhang2018smart} than for the existing scheme \citep{yutaka2019using}.

\begin{figure}[!ht]
    \centering
    \includegraphics[width=8cm, height=5cm]{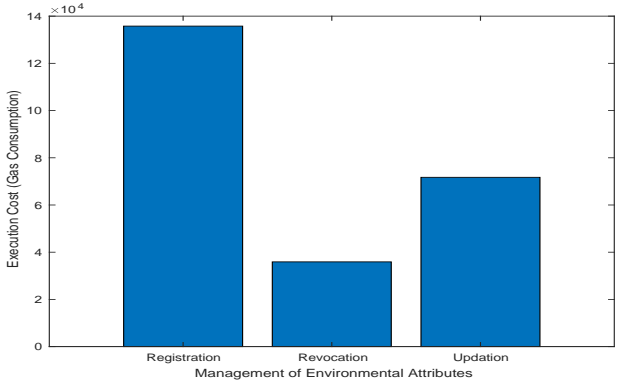}
    \caption{Execution cost for managing environmental attributes}
    \label{executioncostenvironmentattributes}
\end{figure}

Fig. \ref{executioncostenvironmentattributes} depicts the execution cost associated with conducting various management operations on environmental attributes defined in the smart contract $S_{EA}$. As with handling subject and object characteristics, the cost of registering environmental attributes on the Blockchain is greater than the cost of modifying and revoking the attributes. However, because the existing schemes \citep{zhang2018smart} \citep{yutaka2019using} did not employ environmental attributes to define policies, we are only demonstrating the execution cost of managing environmental attributes in our proposed scheme.

\begin{figure}[!ht]
    \centering
    \includegraphics[width=8cm, height=5cm]{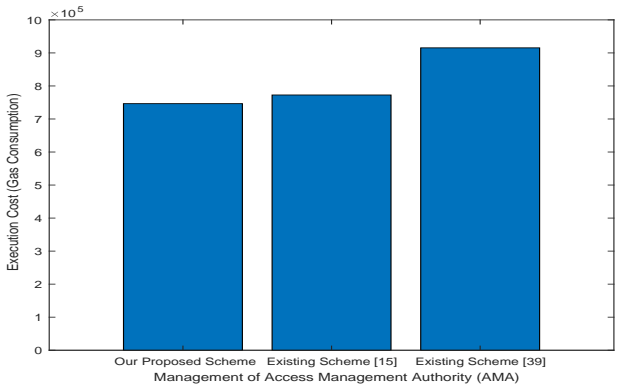}
    \caption{Execution cost for AMA management }
    \label{executioncostAMA}
\end{figure}

Fig. \ref{executioncostAMA} illustrating the execution cost of performing different managing operations on attributes under the AMA component and providing the comparison of our proposed scheme with the existing schemes \citep{zhang2018smart} \citep{yutaka2019using}. We conclude from this comparison that our proposed scheme consumes less Gas for performing the different managing operations on attributes.

\begin{figure}[!ht]
    \centering
    \includegraphics[width=8cm, height=5cm]{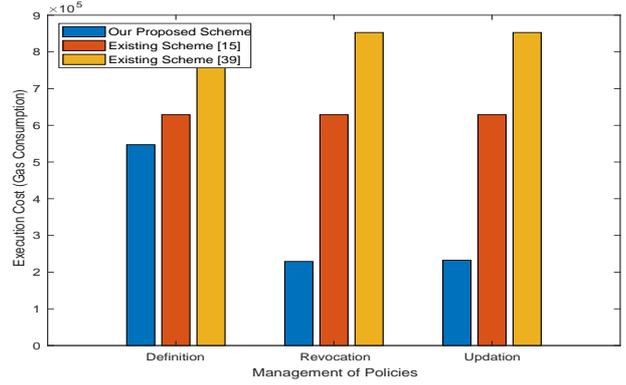}
    \caption{Execution cost for policies management}
    \label{executioncostpolicies}
\end{figure}

Fig.  \ref{executioncostpolicies} depicts the execution cost associated with managing policies on the Blockchain. Similarly to the various management operations performed on attributes, we defined the same managing activities for policies on the Blockchain, including initialise, updating, and revocation. As can be observed, the cost of initialising policies on the Blockchain is substantially more than the cost of modifying or revoking policies. Additionally, the cost of managing policies for the scheme \citep{zhang2018smart}  is greater than that of the scheme \citep{yutaka2019using}. 

\begin{figure}[!ht]
    \centering
    \includegraphics[width=8cm, height=5cm]{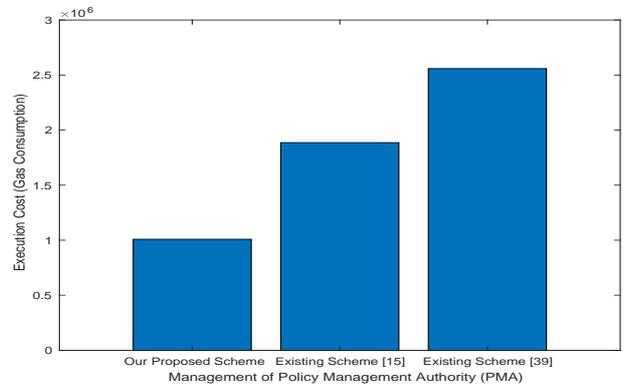}
    \caption{Execution cost for PMA management}
    \label{executioncostPMA}
\end{figure}

Fig. \ref{executioncostPMA} illustrates the execution costs associated with managing policies under our proposed scheme and compares them with the existing schemes \citep{zhang2018smart} and \citep{yutaka2019using}. We conclude from this comparison that our proposed scheme consumes less Gas when performing the various management activities on policies.

\begin{figure}[!ht]
    \centering
    \includegraphics[width=8cm, height=5cm]{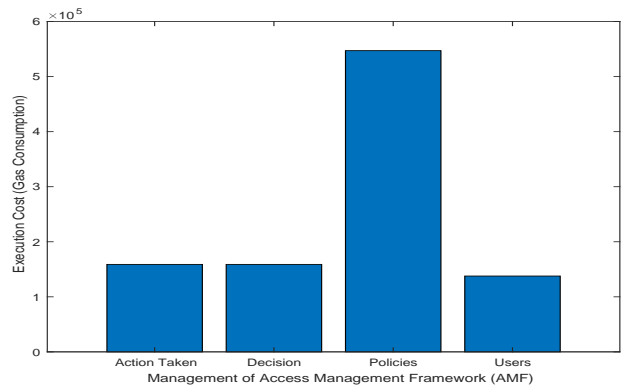}
    \caption{Execution costs for AMF management}
    \label{executioncostAMF}
\end{figure}

Fig. \ref{executioncostAMF} depicts the execution costs associated with the AMF component responsible for managing user requests for access to IoT devices. As can be observed, the cost of executing policies is substantially higher, as the AMF component is responsible for retrieving policies from the Blockchain and comparing them to user requests, determining whether the particular user is permitted to use IoT resources or not. Additionally, the cost of doing various processes such as evaluating the subject's attributes, decisions, and taking action is almost the same

\subsection{Estimated financial cost associated with IoT-based smart home}

\begin{table*}[t]
\small
\centering
\caption{A comparison of the estimated financial costs for our proposed scheme with existing schemes}
\begin{center}
\scriptsize
\begin{tabular}{|c|c|c|c|c|c|c|c|c|}
\hline
\multicolumn{9}{|c|}{\textbf{Our Proposed Scheme}}                                                                                                                                                    \\ \hline
\multicolumn{1}{|c|}{\textbf{Cases}} & \multicolumn{1}{c|}{\textbf{\makecell{Miner \\  Nodes}}} & \multicolumn{1}{c|}{\textbf{\makecell{IoT \\ Devices}}} & \multicolumn{1}{c|}{\textbf{\makecell{Total No. \\ of \\ Devices}}} & \multicolumn{1}{c|}{\textbf{\makecell{No. of \\ Transactions}}} & \multicolumn{1}{c|}{\textbf{\makecell{ Transaction \\ Approx. \\ Cost \\ (In Gas)}}} & \multicolumn{1}{c|}{\textbf{\makecell{Total Gas \\ Consumed}}} &  \multicolumn{1}{c|}{\textbf{\makecell{Gas Cost \\ (ETC)}}} & \multicolumn{1}{c|}{\textbf{\makecell{Price \\ (AUD)}}}   \\ \hline

\multicolumn{1}{|c|}{Case 1} & \multicolumn{1}{c|}{5} & \multicolumn{1}{c|}{15} & \multicolumn{1}{c|}{20} & \multicolumn{1}{c|}{300} & \multicolumn{1}{c|}{2296088} & \multicolumn{1}{c|}{688826400} & \multicolumn{1}{c|}{0.68883}&\multicolumn{1}{c|}{52.7709905}    \\ \hline

\multicolumn{1}{|c|}{Case 2} & \multicolumn{1}{c|}{25} & \multicolumn{1}{c|}{50} & \multicolumn{1}{c|}{75} & \multicolumn{1}{c|}{1125} & \multicolumn{1}{c|}{2296088} & \multicolumn{1}{c|}{2583099000} & \multicolumn{1}{c|}{2.5831}& \multicolumn{1}{c|}{197.891214}   \\ \hline

\multicolumn{1}{|c|}{Case 3} & \multicolumn{1}{c|}{50} & \multicolumn{1}{c|}{150} & \multicolumn{1}{c|}{200} & \multicolumn{1}{c|}{3000} & \multicolumn{1}{c|}{2296088} & \multicolumn{1}{c|}{6888264000} & \multicolumn{1}{c|}{6.88826}& \multicolumn{1}{c|}{527.709905}  \\ \hline

\multicolumn{9}{|c|}{\textbf{Existing Scheme \citep{zhang2018smart}}}                                                                                                                                                    \\ \hline
\multicolumn{1}{|c|}{Case 1} & \multicolumn{1}{c|}{5} & \multicolumn{1}{c|}{15} & \multicolumn{1}{c|}{20} & \multicolumn{1}{c|}{300} & \multicolumn{1}{c|}{1910838} & \multicolumn{1}{c|}{573251400} & \multicolumn{1}{c|}{0.57325}& \multicolumn{1}{c|}{43.9167898}  \\ \hline

\multicolumn{1}{|c|}{Case 2} & \multicolumn{1}{c|}{25} & \multicolumn{1}{c|}{50} & \multicolumn{1}{c|}{75} & \multicolumn{1}{c|}{1125} & \multicolumn{1}{c|}{1910838} & \multicolumn{1}{c|}{2149692750} &\multicolumn{1}{c|}{2.14969} & \multicolumn{1}{c|}{164.687962}   \\ \hline

\multicolumn{1}{|c|}{Case 3} & \multicolumn{1}{c|}{50} & \multicolumn{1}{c|}{150} & \multicolumn{1}{c|}{200} & \multicolumn{1}{c|}{3000} & \multicolumn{1}{c|}{1910838} & \multicolumn{1}{c|}{5732514000} & \multicolumn{1}{c|}{5.73251}&\multicolumn{1}{c|}{439.167898}  \\ \hline

\multicolumn{9}{|c|}{\textbf{Existing Scheme \citep{yutaka2019using}}}                                                                                                                                                    \\ \hline
\multicolumn{1}{|c|}{Case 1} & \multicolumn{1}{c|}{5} & \multicolumn{1}{c|}{15} & \multicolumn{1}{c|}{20} & \multicolumn{1}{c|}{300} & \multicolumn{1}{c|}{2238680} & \multicolumn{1}{c|}{671604000} & \multicolumn{1}{c|}{2518515000}&\multicolumn{1}{c|}{51.4515824}  \\ \hline

\multicolumn{1}{|c|}{Case 2} & \multicolumn{1}{c|}{25} & \multicolumn{1}{c|}{50} & \multicolumn{1}{c|}{75} & \multicolumn{1}{c|}{1125} & \multicolumn{1}{c|}{2238680} & \multicolumn{1}{c|}{2518515000} &\multicolumn{1}{c|}{2.51852}& \multicolumn{1}{c|}{192.943434}  \\ \hline

\multicolumn{1}{|c|}{Case 3} & \multicolumn{1}{c|}{50} & \multicolumn{1}{c|}{150} & \multicolumn{1}{c|}{200} & \multicolumn{1}{c|}{3000} & \multicolumn{1}{c|}{2238680} & \multicolumn{1}{c|}{6716040000} &\multicolumn{1}{c|}{6.71604}& \multicolumn{1}{c|}{513.515824}  \\ \hline

\end{tabular}
\label{tab3}

\end{center}
\end{table*}

Apart from calculating the deployment and execution costs of our proposed authorisation scheme, we also calculated and evaluated the financial costs of deploying it using the IoT-based smart home scenario (as illustrated in the section \ref{casestudy}). In this scenario, we considered different cases in which the number of miner nodes and IoT devices is grown incrementally. Further, in each case, we counted the number of transactions required for performing the different operations, such as registering attributes to their management, initialising policies to their management, and then providing secure access to IoT devices. Further, we estimated the transaction cost, in Gas, for performing different operations. We used Gwei\footnote{1 Ether = 0.000000001 Gwei} as the ether unit since it is the most frequently used gas pricing unit in the Ethereum Blockchain. We calculate the cost of our proposed scheme using Eq. \ref{eq1}, which we refer to informally as our work \citep{hameed2021formally}. A single ETC transaction cost is approximately \$76.61\footnote{1 ETC = \$76.61 (AUD)- Writing a paper on November 16, 2021}. The financial cost equation is provided below.


\begin{equation} \label{eq1}
\begin{split}
\textrm{Financial Cost}  & =  \textrm{No. of Transactions} \times \\ &  \textrm{Transactions cost in Gas} \times \\ &\textrm{Gas cost in ETC} \times \textrm{ETC cost in AUD}
\end{split}
\end{equation}

Table \ref{tab3} presents the detailed estimated total financial cost for deploying our proposed authorisation scheme in terms of the defined smart contracts and the different functions underlying them to give an idea of the total cost associated with the IoT smart home scenario. To perform different functions on the Ethereum Blockchain, we count the number of transactions successfully performed to provide secure authorisation to IoT devices. Next, we calculate the financial costs associated with each case (from Case 1 to Case 3) and then add the costs associated with each case to calculate the overall cost. Financial costs are expressed in Australian dollars (AUD). Further, we provide the financial costs comparison of our proposed scheme with the existing schemes \citep{zhang2018smart} \citep{yutaka2019using}.

\begin{figure}[!ht]
    \centering
    \includegraphics[width=8cm, height=5cm]{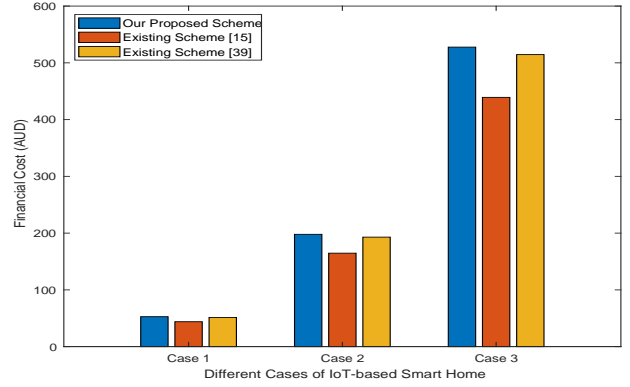}
    \caption{Financial costs associated with deployment of IoT-based smart home }
    \label{Financialcost}
\end{figure}

Fig. \ref{Financialcost} demonstrating the financial costs (in Australian Dollars) associated with deploying our proposed scheme on the Ethereum Blockchain for various scenarios involving varying numbers of miner devices and IoT devices (as described in Table\ref{tab3}). The financial cost of adopting our proposed scheme for the defined IoT-based smart house scenario is slightly higher than the existing schemes \citep{zhang2018smart} and \citep{yutaka2019using}. However, we can confidently state that, in comparison to existing schemes, our proposed scheme introduced the concept of environmental-related attributes to enable dynamic and fine-grained policies and an AMF component for providing secure access to users and enabling auditability for users and their assigned policies. As a result, our scheme's financial costs are almost identical to those of the scheme \citep{zhang2018smart} and slightly higher than those of the scheme \citep{yutaka2019using}.

\section{Conclusion}\label{conclusion}

In this paper, we propose a Blockchain-based secure, decentralised, and flexible authorisation scheme to address the issue of unauthorised access to IoT network devices. We implement the ABAC model using smart contracts, which enables the process of authorising users with secure access to IoT devices to be executed based on dynamic and fine-grained policies stored on the distributed immutable ledger. We designed the ABAC-PMF by utilising smart contracts to manage attributes and policies on the Blockchain. We divided the functionality of ABAC-PMF into two sub-components, namely AMA and PMA, which are responsible for initialisation, storing, and managing attributes (e.g., subject, object, environment), and policies, respectively, on the Blockchain. Furthermore, we designed the AMF to manage user requests to IoT devices based on policies recorded on the Blockchain and to secure the privacy of users and the auditability of user requests to access resources and the policies associated with them. We implemented a prototype as a proof of concept to evaluate the functionality of our proposed scheme on a local Ethereum Blockchain setup. Finally, we evaluate the proposed scheme for an IoT-based smart home scenario in order to determine its applicability in terms of Ethereum gas consumption and financial cost.

\section*{Declaration of Competing Interest}

The authors of this paper declare that they have no known competing financial interests or personal relationships that could have influenced the work reported in this paper.

\section*{Funding}

This research did not receive any specific grant from funding agencies in the public, commercial, or
not-for-profit sectors

 \bibliographystyle{elsarticle-num} 
\bibliography{PaperFile.bib}






\end{document}